\documentclass{aa}
\usepackage[varg]{txfonts}
\usepackage{graphicx}
\usepackage{txfonts}
\usepackage{natbib}
\bibpunct{(}{)}{;}{a}{}{,} 

\makeatletter
\renewcommand*\aa@pageof{, page \thepage{} of \pageref*{LastPage}}
\makeatother

\usepackage{parskip} 

\usepackage{hyperref}
\hypersetup{
    colorlinks = true,
    citecolor ={blue}
}

\usepackage{color}

\usepackage{todonotes} 
\usepackage[]{hyperref} 
\usepackage{tabularx} 
\newcommand\setrow[1]{\gdef\rowmac{#1}#1\ignorespaces} 
\newcommand\clearrow{\global\let\rowmac\relax} 

\begin{document}

   \title{Asteroseismic analysis of variable hot subdwarf stars observed with \emph{\textit{TESS}}}
   \subtitle{I. The mean g-mode period spacings in hot subdwarf B stars\thanks{Based on observations at the La Silla Observatory of the European Southern Observatory for programs number 0103.D-0511 and 0104.D-0514. Based on observations obtained at Las Campanas Observatory under the run code 0KJ21U8U.}
   }
\author{Murat Uzundag\inst{1,2}, Maja Vu\v{c}kovi\'{c}\inst{1}, P\'eter N\'emeth\inst{3,4}, M. Miller Bertolami\inst{5}, Roberto Silvotti\inst{6}, Andrzej S. Baran\inst{7,8,11}, John H. Telting\inst{9,10}, Mike Reed\inst{11}, K. A. Shoaf\inst{11}, Roy H. {\O}stensen\inst{11}, and Sumanta K. Sahoo\inst{7,12} }
  
\institute{Instituto de F\'isica y Astronom\'ia, Universidad de Valpara\'iso, Gran Breta\~na 1111, Playa Ancha, Valpara\'iso 2360102, Chile
\\
\email{murat.uzundag@postgrado.uv.cl}
\and
European Southern Observatory, Alonso de Cordova 3107, Santiago, Chile
\and 
Astronomical Institute of the Czech Academy of Sciences, CZ-251\,65, Ond\v{r}ejov, Czech Republic
\and
Astroserver.org, F\H{o} t\'er 1, 8533 Malomsok, Hungary
\and
Instituto de Astrofísica de La Plata, UNLP-CONICET, La Plata, Paseo del Bosque s/n, B1900FWA, Argentina
\and
INAF-Osservatorio Astrofisico di Torino, strada dell'Osservatorio 20, 10025 Pino Torinese, Italy
\and
ARDASTELLA Research Group, Institute of Physics, Pedagogical University of Cracow, ul. Podchor\c{a}\.zych 2, 30-084 Krak\'ow, Poland
\and
Embry-Riddle Aeronautical University, Department of Physical Science, Daytona Beach, FL\,32114, USA
\and  
Nordic Optical Telescope, Rambla Jos\'e Ana Fern\'andez P'erez 7, 38711 Bre\~na Baja, Spain
\and
Department of Physics and Astronomy, Aarhus University, Ny Munkegade 120, DK-8000 Aarhus C, Denmark
\and
Department of Physics, Astronomy and Materials Science, Missouri State University, 901 S. National, Springfield, MO 65897, USA
\and
Nicolaus Copernicus Astronomical Centre of the Polish Academy of Sciences, ul. Bartycka 18, 00-716 Warsaw, Poland
\\
        }
\date{}


  \abstract
   {We present photometric and spectroscopic analyses of gravity (g-mode) 
   long-period pulsating hot subdwarf B (sdB) stars, also called V1093 Her stars, observed by the \textit{\textit{TESS}} space telescope in both 120\,s short-cadence and 20\,s ultra-short-cadence mode during the survey observation and the extended mission of the southern ecliptic hemisphere.}  
   {We perform a detailed asteroseismic and spectroscopic analysis of five pulsating sdB stars observed with {\it TESS} aiming at the global comparison of the observations with the model predictions based on our stellar evolution computations coupled with the adiabatic pulsation computations.}
   {We process and analyze {\it TESS} observations of long-period pulsating hot subdwarf B stars.
   We perform standard pre-whitening techniques on the datasets to extract the pulsation periods from the {\it TESS} light curves. We apply standard seismic tools for mode identification, including asymptotic period spacings and rotational frequency multiplets.  
   Based on the values obtained from Kolmogorov-Smirnov and Inverse Variance tests, we search for a constant period spacing for dipole ($l = 1$) and quadrupole ($l = 2$) modes.
   We calculate the mean period spacing for $l = 1$ and $l = 2$ modes and estimate the errors by means of a statistical resampling analysis. For all stars, atmospheric parameters were derived by fitting synthetic spectra to the newly obtained low-resolution spectra.
   We have computed stellar evolution models using {\tt LPCODE} stellar evolution code, and computed $l = 1$ g-mode frequencies with the adiabatic non-radial pulsation code {\tt LP-PUL}.
   Derived observational mean period spacings are then compared to the mean period spacings from detailed stellar evolution computations coupled with the adiabatic pulsation computations of g-modes.}
   {We detect 73 frequencies, most of which are identified as dipole and quadrupole g-modes with periods spanning from $\sim 3\,000$\,s to $\sim 14\,500$\,s. 
   The derived mean period spacing of dipole modes is concentrated in a narrow region ranging from 251\,s to 256\,s, while the mean period spacing for quadrupole modes spans from 145\,s to 154\,s. 
   The atmospheric parameters derived from spectroscopic data are typical of long-period pulsating sdB stars with the effective temperature ranging from 23\,700\,K to 27\,600\,K and surface gravity spanning from 5.3\,dex to 5.5\,dex. In agreement with the expectations from theoretical arguments and previous asteroseismological works, we find that the mean period spacings obtained for models with small convective cores, as predicted by a pure Schwarzschild criterion, are incompatible with the observations. We find that models with a standard/modest convective boundary mixing at the boundary of the convective core are in better agreement with the observed mean period spacings and are therefore more realistic.}
   {Using high-quality space-based photometry collected by the {\it TESS} mission coupled with low-resolution spectroscopy from the ground, we have provided a global comparison of the observations with model predictions by means of a robust indicator such as the mean period spacing.
   All five objects that we analyze in this work show a remarkable homogeneity in both seismic and spectroscopic properties.}


    \titlerunning{The mean g-mode period spacings in pulsating hot subdwarf B stars} 
	\authorrunning{Murat Uzundag et al.}
   \maketitle

\section{Introduction}

Hot subdwarf stars (sdB) are core-helium burning stars with a very thin hydrogen (H) envelope ($M_{\rm env}$ $<$ 0.01 $M_{\odot}$), and a mass close to the core-helium (He)-flash mass $\sim$0.47 $M_{\odot}$. The sdB stars are evolved compact ($\log{g}$ = 5.0 - 6.2\,dex) and hot ($T_{\rm eff}$ = 20\,000 - 40\,000\,K) objects with radii between 0.15 $R_{\odot}$ and 0.35 $R_{\odot}$, located on the so-called extreme horizontal branch \citep[EHB; see][for a review]{heber2016}.
They have experienced extreme mass-loss mostly due to binary interactions at the end of the red giant phase, where they lost almost the entire H-rich envelope, leaving a He burning core with an envelope too thin to sustain H-shell burning. 
Hot subdwarf B stars will spend about $10^{8}$\,yr burning He in their cores. Once the He has been exhausted in their core, they will start burning He in a shell surrounding a carbon/oxygen (C/O) core as subdwarf O (sdO) stars, and eventually they will end their lives as 
white dwarfs \citep{heber2016}.

One of the major progresses in our understanding of sdB stars was initiated by \citet{kilkenny1997}, who discovered rapid pulsations in hot sdBs known as V361 Hya stars (often referred to as short-period sdBV stars). The V361 Hya stars show multiperiodic pulsations with periods spanning from 60\,s to 800\,s. In this region, the pulsational modes correspond to low-degree, low-order pressure p-modes with photometric amplitudes up to a few per cent of their mean brightness \citep{reed2007, ostensen2010, green2011}. 
These modes are excited by a classical $\kappa$-mechanism due to the accumulation of the iron group elements (mostly iron itself), in the $Z$-bump region \citep{charpinet1996,charpinet1997}. The authors also showed that radiative levitation is a key physical process to enhance the abundances of iron group elements in order to be able to excite the pulsational modes.  
The p-mode sdB pulsators are found in a temperature range between 28\,000\,K and 35\,000\,K and with the surface gravity $\log{g}$ in the interval 5-6 dex. 
Later on, the long-period sdB pulsators known as V1093 Her stars were discovered by \citet{green2003}, which show brightness variations with periods of up to a few hours and have amplitudes smaller than 0.1 per cent of their mean brightness \citep{reed2011}.
 The oscillation frequencies are associated with low degree ($l <$ 3) medium- to high-order (10 $<$ $n$ $<$ 60) g-modes, which are driven by the same mechanism \citep{fontaine2003,charpinet2011}. 
The g-mode sdB pulsators are somewhat cooler with temperatures ranging from 22\,000\,K to 30\,000\,K and $\log{g}$ from 5\,dex to 5.5\,dex.
Between the two described families of pulsating sdB stars, some "hybrid" sdB pulsators, which simultaneosuly show g- and p-modes, have been found. 
These hybrid sdB pulsators are located in the middle of the region of the HR diagram between p- and g-mode sdB pulsators  \citep[Figure 5]{green2011}.  
These objects are of particular importance as they enable us to study both the core structure and the outer layers of the sdBVs via asteroseismology. 
A few examples were found from the ground \citep{oreiro2005,baran2005,schuh2006,lutz2008}.

During the nominal \textit{Kepler} mission \citep{borucki2010}, 18 pulsating subdwarf B stars were monitored in 1\,min short-cadence mode. The majority of the stars (16) were found to be long-period g-mode pulsators, while just two of them were short period p-mode pulsators \citep{ostensen2010,ostensen2011,baran2011,reed2011,kawaler2010}. 
Additionally, 3 known sdBs stars in the old open cluster NGC\,6791 were found to pulsate \citep{pablo2011, reed2012}. The temperature of these sdBs range from 21\,500\,K to 37\,000\, K with a median of 27\,400\,K and surface gravities  between $\log{g}$ of 4.67\,dex and 5.82\,dex, with a median of 5.42\,dex. 
In 2013, the \textit{Kepler} mission was re-initiated after the second reaction-wheel failure and it continued as the K2 mission observing along the ecliptic \citep{haas2014}. During the K2 mission more than 25 sdBs have been found to pulsate and the analyses are still ongoing.  Thus far 18 of these sdBVs have been published together with their atmospheric parameters \citep{reedk22016, kernk22017, bachulskik22016, reedk22019, barank22019, silvottik22019, reedk22020, ostensen2020}. These stars have $T_{\rm eff}$ in between 22\,300\,K and 37\,000\,K and $\log{g}$ is from 5.2\,dex to 5.7\,dex \citep[for a review]{Reed2018}.  

For many sdBVs observed with \textit{Kepler} the asymptotic period sequences for g-mode pulsations have been successfully applied, especially for dipole ($l = 1$) and quadrupole ($l = 2$) modes, as more than 60\% of the periodicities are associated with these modes \citep{Reed2018}. 
The asymptotic approximation can be perfectly applied for homogeneous stars. 
The period separation of g-modes becomes approximately constant for high radial orders. 
This is called the asymptotic regime of pulsations, and it is masterfully documented in \citet{Tassoul1980}. 
It is important to note that the asymptotic g-mode theory is strictly valid for completely radiative and chemically homogeneous stars.
However, sdB stars are stratified and diffusion processes (gravitational settling and radiative levitation) contribute significantly to compositional discontinuities, which disturb the pulsational modes and could break the sequences of periods with constant spacing. 
This effect has been found in several g-mode dominated sdBV stars. 
Furthermore, when the compositional discontinuities become stronger in transition zones, some modes are trapped, which was also detected for a few sdBV stars observed with \textit{Kepler} \citep{ostensen2014,uzundag2017,baran2017,kern2018}.
Mode trapping is characterized by strong departures from a constant period spacing.
Trapped modes can be quite useful to provide a test of stellar evolution models, and offer a unique opportunity to determine mixing processes due to convective overshooting beyond the boundary of the helium burning core \citep{Ghasemi2017}. 

Another important asteroseismic tool, rotational multiplets, became available for sdB stars thanks to the long baseline of \textit{Kepler} data \citep{baran2012a}.
During the nominal mission of $\textit{Kepler}$, the rotation periods of sdBs have been found to range from 10\,d to 100\,d \citep{reed2014}. 
For the short-period sdB systems with WD companions (P$\rm orb$ $<$ 15\,d) the rotational periods of the sdBs were found to be in the range from 28\,d to 50\,d \citep[references therein]{barank22019}. 
For the short-period sdB binary systems with M-dwarf companions (P$\rm orb$ $<$ 0.8\,d) rotational periods in the range between 7 and 40 days have been identified \citep{barank22019}.
Moreover, detecting rotational multiplets in both g- and p-modes in hybrid sdB pulsators is of special importance as it provides a way to determine the rotation of both the core and the envelope of these stars. 
Also, rotational splittings allow us to assign a harmonic degree to a pulsation mode.
 These remarkable stars have been found to be either solid-body rotators \citep{baran2012a, kernk22017} or radially differential rotators \citep{foster2015, baran2017}. The rotation properties of evolved stars are further discussed in a recent review of \citet{Charpinet2018}.
  
The Transiting Exoplanet Survey Satellite (\textit{TESS}) was launched successfully on 18 April 2018. The primary goal of this mission is to discover exoplanets around nearby and bright stars by means of the transit method \citep{Ricker2014}.  
The spacecraft has four identical 100\,mm aperture cameras and is in a highly eccentric lunar-resonance Earth orbit. The orbit allows the telescope to observe the targets during $\sim$27 days continuously covering a huge area in the sky ($24^{o}$\,x\,$96^{o}$). 
During the first year, \textit{TESS} surveyed 13 sectors in the southern hemisphere with both 2-min short cadence (SC) and 30-min long cadence (LC). Results from LC observations of the first year have been reported by \citet{Sahoo2020}.

During the first year, \textit{TESS} observed 1702 compact objects including hot subdwarfs, pre-white dwarfs and white dwarfs with 2-min cadence. 
The first results regarding asteroseismic analysis of hot sdB pulsating stars observed by \textit{TESS} have been reported in three papers \citep{charpinet2019,reed2020,saho2020}. 
In this paper, we analyze 5 pulsating sdB stars, which were observed in a single sector in SC mode by \textit{TESS} during the survey phase of the southern ecliptic hemisphere.
For each of these \textit{TESS} targets we have obtained low-resolution spectra and fitted model atmospheres in order to derive their fundamental atmospheric parameters. We present the details of spectroscopic and photometric observations as well as the main characteristics of the studied sdBVs in Sect.~\ref{sect:obs}. 
We discuss the analysis of the \textit{TESS} data in Sect.~\ref{sect:TESS_data} and give details on the frequency analysis along with detailed seismic mode identification.
In Sect.~\ref{sect:spect_data} and \ref{sect:spect_analysis}, we analyzed the spectroscopic data and derive atmospheric parameters for each star by fitting synthetic spectra to the newly obtained low-resolution spectra.
We calculate asteroseismic models in Sect.~\ref{sect:model_seismology} and compare them with the observations. 
Finally, we summarize our findings in Sect.~\ref{sect:summary}.

\section{Observations}
\label{sect:obs}

\subsection{Photometric observations --- \textit{TESS}}
 
The \textit{TESS} mission Cycle 1, covering most of the southern hemisphere, started on 25 July 2018 and ended on 18 July 2019. During this time a total of 13 sectors was observed, where each sector covered $\sim$27 days of continuous observations with a 2-min short-cadence (SC) mode. During Cycle 1, \textit{TESS} observed 806 subdwarf (sd) candidates including sdB stars, sdO stars and their He-rich counterparts. Among them, we have found several rich oscillators including short- and long-period sdB pulsators. By "rich", we mean that there are sufficient pulsation frequencies present for asteroseismic methods (e.g. rotational multiplets and/or asymptotic period spacing) to be applicable.
 Thus far, 5 rich long-period sdB pulsators have been reported.
Four stars have been analyzed by \citet{saho2020} and \citet{reed2020} by applying the asymptotic period spacing. For one target, \citet{charpinet2019} have produced a detailed model by best-matching all the observed frequencies with those computed from models.
In this paper, we have concentrated on 5 other long-period pulsating sdB stars, TIC\,260795163, TIC\,080290366, TIC\,020448010, TIC\,138707823 and TIC\,415339307 that were observed by \textit{TESS}.
 During Cycle 2, the second year of the primary mission, \textit{TESS} observed the northern hemisphere, sectors 14-26, after which it re-observed the southern hemisphere in what is referred to as extended mission.
During the extended \textit{TESS} mission, three of the selected stars (TIC\,260795163, TIC\,080290366 and TIC\,138707823) have been observed with 20-sec ultra-short-cadence (USC) mode. While TIC\,080290366 and TIC\,138707823 were observed in USC mode during only one sector (29), TIC\,260795163 was observed in USC mode during two consecutive sectors (27 and 28).

Among the stars analyzed in this paper, the only star for which a photometric variability was discovered before \textit{TESS} is TIC\,080290366 \citep{koen2010}. The remaining four V1093 Her stars are new discoveries. The details of the photometric \textit{TESS} observations are summarized in Table \ref{table1}, where we also give the literature name of the targets, which are taken from SIMBAD \footnote{http://simbad.u-strasbg.fr/simbad/}, the \textit{TESS} Input Catalog (TIC) number, right ascension, declination and $T$-magnitude along with their corresponding observed sectors.
Using available magnitude values from the literature, we have calculated \textit{TESS} magnitude of all targets as described by \citet{stassun2018} using the tool of \textit{ticgen}\footnote{https://github.com/tessgi/ticgen}.  

\subsection{Spectroscopic observations}

The spectroscopic follow-up observations of the sdB pulsators analyzed in this paper were 
obtained with two instruments, the Boller and Chivens (B\&C) spectrograph mounted at the 2.5-meter (100-inch) Ir\'ene du Pont telescope at Las Campanas Observatory in Chile\footnote{For a description of instrumentation, see: \url{http://www.lco.cl/?epkb_post_type_1=boller-and-chivens-specs}}, and the European Southern Observatory (ESO) Faint Object Spectrograph and Camera (v.2) (EFOSC2) \citep{buzzoni1984} mounted at the Nasmyth B focus of the New Technology Telescope (NTT) at La Silla Observatory in Chile.

We obtained low-resolution spectra in order to calculate the atmospheric parameters, such as effective temperature $T_{\rm eff}$, surface gravity $\log{g}$ and He abundance 
Even though the atmospheric parameters for TIC\,260795163, TIC\,080290366, TIC\,020448010, TIC\,138707823 and TIC\,415339307 are available in the literature \citep{heber1984,Heber1986,kilkenny1995,killkennyheber1988,nemeth2012, Lei18}, we reobserved them in order to ensure homogeneity in our analysis.
The B\&C spectra were obtained using the 600\,lines/mm grating corresponding to the central wavelength of 5\,000\,\AA, covering a wavelength range from 3\,427 to 6\,573\,\AA. We used a 1 arcsec slit, which provided a resolution of 3.1\,\AA. 
Depending on the brightness of the targets, the exposure times were between 300\,s and 480\,s, which was enough to obtain an optimal signal-to-noise ratio (S/N) to measure $T_{\rm eff}$ and $\log{g}$ with 5\% precision.
For the EFOSC2 setup, we used grism \#7 and a 1 arcsec slit and the exposure times were between 200\,s and 300\,s. 
This setup provided a wavelength coverage from 3270 to 5240 \AA with a S/N of about 150. 
TIC\,260795163, TIC\,080290366 and TIC\,138707823 were observed with 2x2 binning mode at a resolution of 5.4\,\AA, while TIC\,020448010 was observed using 1x2 binning, such that the spectral resolution slightly improved to 5.2\,\AA.
The details of the spectroscopic observations are given in Table \ref{tablespec1} including, instrument, date, exposure time, resolution and S/N ratio at 4\,200\,\AA. 

\begin{table*}
\caption{Five sdB stars studied in this work, including the \textit{TESS} input catalog number, the name of the star from the Simbad database, right ascension, declination, \textit{TESS} magnitude and \textit{TESS} observed sectors (including ultra-short cadence (USC) observations) and distances from $Gaia$, respectively.}
\begin{tabular}{clcccccc}
\hline \hline
TIC & Name & RA(J2000) & Dec(J2000) & $T_{\rm mag}$ &  Observed Sectors (USC) & Distance (pc)\\
\hline
         
260795163 & EC23073-6905            & 23:10:35.5 & -68:49:30.2  & 11.73 & 1 (27-28) & 499.6 $\pm$ 14 \\
080290366 & JL194                   & 00:31:41.6 & -47:25:20.1  & 11.85 & 2 (29) & 502.1 $\pm$ 13 \\
020448010 & GALEXJ11143-2421        & 11:14:22.0 & -24:21:29.0  & 12.18 & 9 & 509.1 $\pm$ 14 \\
138707823 & FB1                     & 00:03:22.1 & -23:38:58.0  & 12.70 & 2 (29) & 695.9 $\pm$ 28 \\
415339307 & HS0352+1019             & 03:55:14.3 &  +10:28:12.6 & 14.24 & 5 & 771.3 $\pm$ 33 \\

\hline 
\label{table1}
\end{tabular}
\end{table*}

\begin{table*}
\setlength{\tabcolsep}{2.2pt}
\renewcommand{\arraystretch}{1.1}
\centering
\caption{Observing log of the spectroscopic data obtained for the 5 sdB stars studied in this work. Columns correspond to the {\it TESS} input catalog number, the spectrograph that was used for observations (with corresponding number of spectra obtained), the observing date, the exposure times, the spectroscopic data resolution and signal-to-noise level at 4200\,\AA, respectively.}
\begin{tabular}{cccccc}
\hline \hline
TIC &  Spectrograph & Date & $t_{exp} (s)$ & Resolution ($\Delta \lambda$ (\AA)) & S/N (@4200\,\AA)\\
\hline
         
260795163   & B\&C (2) - EFOSC2 (1)  & 23 Aug, 30 Oct 2019,  11 Jan 2020 & 420 - 300 - 200  &   3.1 - 5.4    &  120 - 80  \\
080290366   & B\&C (2) - EFOSC2 (1)  & 23 Aug, 30 Oct 11, Jan 2020 & 450 - 240 - 200  &   3.1 - 5.4    & 100 - 150   \\
020448010   & EFOSC2 (2)             & 11 Jan 2020 & 240             &   5.2          &  150  \\
138707823   & B\&C (1) - EFOSC2 (1)  & 22, 24 Jan 2020 & 480 - 300       &   3.1 - 5.4    &  100 - 150    \\
415339307   & B\&C (1)   & 20 Aug 2020  &  420      &  3.1         &  70   \\

\hline 
\label{tablespec1}
\end{tabular}
\end{table*}

\subsection{The targets}
\begin{itemize}

\item TIC\,260795163 (EC\,23073-6905) was discovered during Edinburgh-Cape  survey-II \citep{kilkenny1995} and was classified as an sdB star with low-dispersion spectrograms  and \textit{UBV} photometry. \citet{kilkenny1995} derived $T_{\rm eff} \sim 27000$ K and $\log{g}\sim5$ dex, respectively. They reported a radial velocity variation of about $\pm$ 26\,km/s. Afterwards, \cite{magee1998} and \cite{copperwheat2011} measured the radial velocity of the star and they did not find a significant variation. 
The {\it Gaia} DR2 parallax and corresponding distance for this object are $\pi= 2.002 \pm0.058$\,mas and $d= 499.6\pm 14.4$\,pc. \\

\item TIC\,080290366 (alias JL\,194, EC\,00292-4741, CD-48 106) is a well-known relatively bright hot subdwarf star with $T$-band magnitude of 11.85.
The star was observed several times and can be found in many surveys, including \citet{hill1966, jaidee1969, kilkenny1975, wagner1980, kilkenny1988, Kilkenny2016}.
The atmospheric parameters of TIC\,080290366 have been derived by \citet{heber1984} and also given by \citet{killkennyheber1988}. 
The authors showed that the effective temperature and surface gravity of TIC\,080290366 are $T_{\rm eff}= 25\,200$\,K and $\log g=5.20$ dex.
The evolutionary status of TIC\,080290366 was discussed by \citep{newell1973}. 
The presence of a potential weak magnetic field was investigated by \citet{mathys2012}, however, the detection limit was not enough to be conclusive. The parallax and corresponding distance for this star extracted from {\it Gaia} DR2 are $\pi= 1.992 \pm 0.05$\,mas and $d= 502.1 \pm 12.6$\,pc.\\

\item TIC\,020448010 (EC 11119-2405) was discovered during the Edinburgh-Cape Blue Object Survey as a hot sdB star with $V$-band magnitude of 12.72 \citep{kilkenny1997}. 
The atmospheric parameters were obtained by \citet{nemeth2012}. 
The authors found $T_{\rm eff}=23\,430\pm900$\,K, $\log{g}=5.29\pm0.15$ dex and a low surface He abundance of $\log({\rm He/H})=-2.52\pm0.25$.
\citet{kawka2015} included the target in their survey for hot subdwarf binaries, but did not detect significant velocity variations. 
From {\it Gaia} DR2, the parallax and distance of this object are $\pi= 1.964 \pm 0.055$\,mas and $d= 509.1 \pm 14.2 $\,pc, respectively. \\

\item The discovery of TIC\,138707823 (alias EC\,00008-2355, FB\,1, Ton\,S\,135, PHL\,2580, MCT\,0000-2355) was led by \citet{Haro1962} who searched for faint blue stars in the region near the south galactic pole. 
TIC\,138707823 was confirmed by \citet{Lamontagne2000} as an sdB star. 
Estimation of atmospheric parameters of TIC\,138707823 have been given in several papers \citep{Greenstein1974,kilkenny1977,killkennyheber1988} and ranged from effective temperature of 23\,000 to 27\,000 K and  surface gravity of 5.4 to 5.6 dex.  
\citet{Heber1986} measured $T_{\rm eff}$ and $\log{g}$ as 25\,600 $\pm$ 1\,250\,K and 5.60 $\pm$ 0.20\,dex, respectively.
\citet{Edelmann2005} found that TIC\,138707823 is a binary system comprising of an sdB and a main sequence (MS) or a white dwarf (WD) companion with an orbital period of P$_{\rm orb}$ = 4.122 $\pm$ 0.008\,d. 
\citet{Geier2012} calculated the atmospheric parameters of TIC\,138707823 from UVES spectroscopy and found $T_{\rm eff}$ = 27\,600 $\pm$ 500\,K and $\log{g}$ = 5.43 $\pm$ 0.05\,dex. 
From {\it Gaia} DR2, the parallax and distance of TIC\,138707823 are $\pi= 1.437 \pm 0.057$\,mas and $d=695.9 \pm 27.7$\,pc, respectively.\\

\item TIC\,415339307 (HS\,0352+1019) was included in the KISO Survey and Hamburg Quasar Survey \citep{wegner1993,edelmann2003}. 
The atmospheric parameters were determined by \citet{edelmann2003}, who found $T_{\rm eff}$ = 24\,900 $\pm$ 600\,K, $\log{g}$ = 5.34 $\pm$ 0.1\,dex and $\log{(n_{\rm He}/n_{\rm H})}$ = -2.7 $\pm$ 0.2.
Recently, \citet{Lei18} has derived $26\,340\pm150$\,K, $5.33\pm0.01$\,dex and $\log{(n_{\rm He}/n_{\rm H})} = -2.68\pm0.06$.
From {\it Gaia} DR2, the parallax and distance of TIC\,415339307 are $\pi= 1.296 \pm 0.056$\,mas and $d=771.36 \pm 33.04$\,pc, respectively.\\

\end{itemize}

         


\section{Analysis of \textit{TESS} data}
\label{sect:TESS_data}

We analyzed {\it TESS} observations using the SC mode, which samples every 2-minutes allowing us to analyze the frequency range up to the Nyquist frequency at about 4\,167 $\mu$Hz. 
Given that the {\it TESS} USC data, with 20-sec sampling time, became available recently for 3 of the analyzed stars, we have included the analysis of the USC data when available.
The Nyquist frequency of the USC data is at about 25\,000 $\mu$Hz which permits to analyse the short period range of the pulsation spectra.   
The light curves were processed by the Science Processing Operations Center (SPOC) pipeline \citep{Jenkins2016}, which is based on the Kepler Mission science pipeline and made available by the NASA Ames SPOC center and at the MAST archive\footnote{http://archive.stsci.edu}. 

We first downloaded the target pixel file (TPF) of interest from the MAST archive, which is maintained by the Lightkurve Collaboration; \citet{lightkurve2020}. 
The TPFs include an 11x11 postage stamp of pixels from one of the four CCDs per camera that the target is located on. The TPFs are examined to determine the amount of crowding and other potential bright sources near the target. 
We used TPFs to optimize the aperture if it was needed. In most of the cases we have used the pipeline aperture as it gave the most optimal result with respect to signal-to-noise ratio. 
Given that the pixel size of \textit{TESS} is huge, 21 arcsec, we need to pay special attention to the possible contamination. The contamination factor is indicated with the keyword $\tt{CROWDSAP}$, which gives the ratio of the target flux to the total flux in the \textit{TESS} aperture. For the each target, we have checked the contamination by looking at $\tt{CROWDSAP}$ parameter which is listed in Table \ref{Table FT}. 
For the 3 stars (TIC\,260795163, TIC\,080290366 and TIC\,138707823) that were also observed during the extended mission we give the relevant parameters in parenthesis in Table \ref{Table FT}.
For 4 targets, TIC\,080290366, TIC\,020448010, TIC\,138707823 and TIC\,415339307, the $\tt{CROWDSAP}$ value is higher than 0.9, which implies that less than 10\% of the total flux originally measured in the \textit{TESS} aperture comes from other unresolved sources. For TIC\,260795163, the $\tt{CROWDSAP}$ is smaller than 0.7 (for SC mode observations) and 
almost 40\% of the flux comes from the other background sources. 
 Since the difference in magnitude between the target and nearby object is more than 4, we can safely conclude that the flux variations indeed come from the sdB star. 
For the extended mission, the $\tt{CROWDSAP}$ is much better for TIC\,260795163 with 0.81 and for TIC\,080290366 and TIC\,138707823 it is the same as for the SC data, i.e., bigger than 0.9.  

We have generated the light curves by integrating the fluxes ('PDCSAP FLUX'\footnote{The pre-search data conditioning (PDC) flux, which corrects the simple aperture photometry (SAP) to remove instrumental trends.}) within the aperture mask as a function of time in barycentric corrected Julian days ('BJD - 2457000').
After that, we removed the outliers that vary significantly from the local standard deviation ($\sigma$) by applying a running $5\sigma$ clipping mask.
Then, we detrended the light curves to remove any additional low-frequency systematics that may be present in the data. To do this, we applied Savitzky–Golay filter with a three-day window length computed with the Python package {\sc lightkurve}. Detrending with this method suppressed any spurious frequencies below 1 $\mu$Hz, which is the typical region where spurious frequencies are seen in \textit{TESS} data.
We have also examined the light curves before applying the low-frequency fitting in order to search for any potential binary signals that might be affected by detrending. However, we did not find binary signatures in the FT of any of the 5 stars. The fluxes were then normalized and transformed to amplitudes in parts-per-thousand (ppt) unit (($\Delta I$/$I-1)\times 1000$). 
For TIC\,260795163, which was observed during two consecutive sectors in USC mode, we have combined the light curves. 

\subsection{Frequency analysis}
\label{sect:freq_analysis}

The Fourier transforms (FT) of the light curves were computed to examine the periodicities present in the data, aiming at identifying the frequency of all pulsation modes, along with their amplitude and phase.

We adopt a rather conservative detection threshold of $0.1$ per cent false alarm probability (FAP), which means that if the amplitude reaches this limit, there is a $0.1$ percent chance that it is just result of noise fluctuations.
We calculated the $0.1\%$ FAP threshold following the method described in \citet{kepler1993}.

The temporal resolution of the data is around 0.6 $\mu$Hz (1.5/T, where T is the data length which is between 24 and 27 d). 
In Table \ref{Table FT}, we listed all relevant information regarding the Fourier transform including number of data points and $0.1\%$ FAP level of each dataset.

For all the peaks that are above the accepted threshold and up to the frequency resolution of the particular dataset, we have performed a nonlinear least square (NLLS) fit in the form of $A_i \sin(\omega_i\ t + \phi_i)$,
with $\omega=2\pi/P$, where $P$ is the period. In this way, we determined the values of frequency (period), phase and amplitude corresponding to each periodicity. 
Using the parameters of NLLS fit, we have prewhitened the light curves until no signal above the $0.1\%$ FAP level was left in the FT of each star unless there were unresolved peaks. For all frequencies that still had some signal left above the threshold after prewhitening we carefully checked if there was a close-by frequency within the frequency resolution and in such case only the highest amplitude frequency was fitted and prewhitened, as shown in Fig. \ref{fig:FT_260_R} and \ref{fig:FT_08_R}.
All prewhitened frequencies for each of the 5 stars are given in Table \ref{260795163}, \ref{080290366}, \ref{020448010}, \ref{138707823} and \ref{415339307}, showing frequencies (periods) and amplitudes with their corresponding errors and the S/N ratio.
The Fourier transforms of the prewhitened light curves of all 5 analysed stars are shown in Figures \ref{fig:260795163}, \ref{fig:080290366}, \ref{fig:020448010}, \ref{fig:138707823} and \ref{fig:415339307}. 

\begin{figure}
    \centering    
    \includegraphics[clip,width=1.\linewidth]{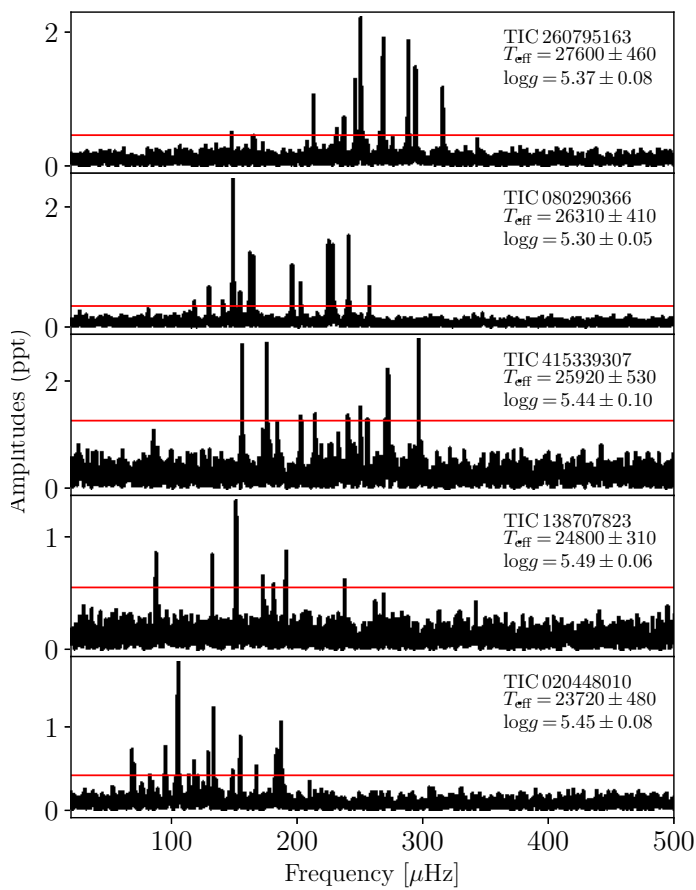} 
 \caption{Fourier transform  of all 5 sdB stars observed in single SC sectors, concentrating on on the g-mode region of the frequency spectrum. 
 The panels are sorted with decreasing effective temperature from top to bottom. In each panel we give the {\it TESS} input catalog number, effective temperature (in Kelvin) and surface gravity (in dex), respectively. The horizontal red lines correspond to $0.1\%$ FAP confidence level.}
    \label{fig:freq-spectra}
\end{figure}

For all 5 stars analyzed in this paper, a total of 73 frequencies were extracted from their light curves. The detected frequencies are distributed in a narrow region between 68 $\mu$Hz and 315 $\mu$Hz.  
This corresponds to the g-mode region seen in V1093\,Her type sdB pulsators \citep[e.g.][]{Reed2018}.
The amplitude spectra of all 5 stars are shown in Fig. \ref{fig:freq-spectra}, where we also give atmospheric parameters derived in Section \ref{sect:spect_data} for each star.
It has been recently reported by \citet{reed2020} that there is a correlation between effective temperature and the frequency of the highest amplitude of g-modes detected in V1093\,Her type sdB pulsators observed by \textit{Kepler} and K2. 
As can be seen in Fig. \ref{fig:freq-spectra}, the five g-mode sdB pulsators analyzed in this paper do not deviate from this finding.

\begin{table*}
\caption{The parameters of Fourier Transforms of g-mode sdB pulsators from this work.
  Columns 
  1 to 7 correspond to the {\it TESS} input catalog number, $Gaia$ magnitudes, duration of each dataset, the $\tt{CROWDSAP}$ keyword, the number of data points, data resolution and the significance level, respectively. 
  In parenthesis, the details from the extended mission observations are given. For more details, see the text.}
\begin{tabular}{ccccccccc}
\hline \hline
TIC &  $G_{mag}$ & Observations [d]  &\tt{CROWDSAP}  & N.data & Resolution [$\mu$Hz]  & $0.1\%$ FAP \\
\hline                   
260795163    & 12.56     & 27.88 (49.39) &  0.61 (0.81)  & 18099 (192385)  & 0.623 (0.352)  & 0.467 (0.278) \\
080290366    & 12.38     & 27.40 (24.25)    &  0.93 (0.94)   & 18312 (88731)  & 0.634 (0.716)  & 0.36 (0.325) \\
020448010    & 12.77     & 24.20      &  0.96   & 15946   & 0.717  & 0.429 \\ 
138707823    & 13.27     & 27.40 (23.85)   &  0.99 (0.99)  & 18317 (85563)  & 0.634 (0.728)  & 0.563 (0.66) \\
415339307    & 14.15     & 25.99 &  0.95   & 17660   & 0.668   & 1.270 \\

\hline 
\label{Table FT}
\end{tabular}
\end{table*}

{\bf TIC\,260795163} ($T_{mag} = 11.73$) was observed in SC during sector 1 between July 25 and August 22 2018. 
The observations yielded 18\,099 data points with the temporal resolution of 0.62\,$\mu$Hz (1.5/T, where T is 27.87\,d). 
From sector 1 data, we detected 12 frequencies above $0.1\%$ FAP confidence level, which corresponds to 0.467\,ppt. 
The median of noise level including the entire FT is equal to 0.09\,ppt. 
The signal-to-noise level of detected frequencies ranges from 4.75 to 30.87.

TIC\,260795163 was also observed during two consecutive sectors of the extended mission. 
These observations started on 04 July 2020 and ended on 26 August 2020. 
From this 49 day long dataset, we calculate the FT of the USC data up to Nyquist frequency of 25\,000 $\mu$Hz.  
The frequency resolution of USC data is 0.35 $\mu$Hz. 
The average noise level of the entire FT of the USC dataset is 0.055 ppt and the $0.1\%$ FAP threshold level is 0.26 ppt. 
All frequencies with amplitudes above this threshold are concentrated in a narrow region between 100 and 320 $\mu$Hz, which is almost identical to what we detected from the SC observations. 
Beyond 320 $\mu$Hz, there is no peak detected above the 0.1 \% FAP threshold up to the  Nyquist frequency. 
We find only one peak reaching the 4.5\,$\sigma$ level at 23\,975.8 $\mu$Hz. However, this frequency seems too high to be excited in an sdB pulsator.

\begin{figure}
    \includegraphics[width=0.5\textwidth]{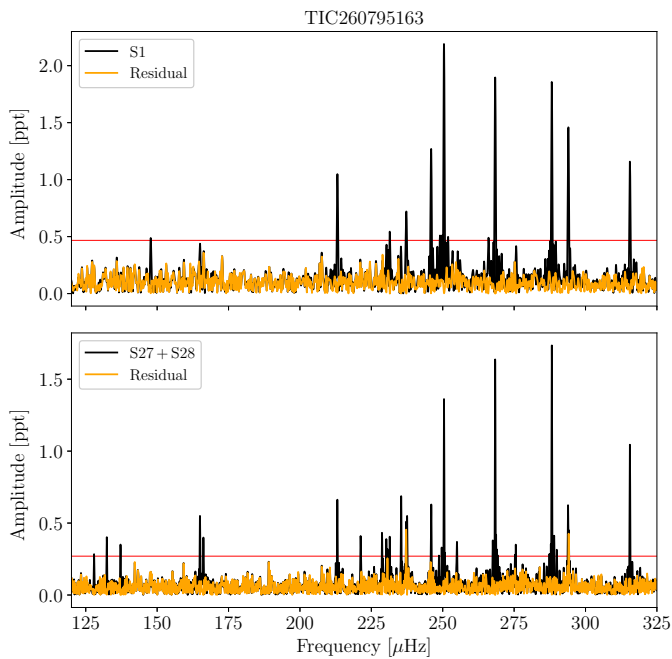} 
 \caption{{\sc Top:} Fourier transform of sector 1 of TIC\,260795163. The horizontal red line indicates the 0.1\% FAP level. The orange line is the residual after extraction of the signals.
 {\sc Bottom:}  Fourier transform of sector 27 and 28 of TIC\,260795163. The horizontal red line indicates the 0.1\% FAP level. The orange line is the FT of the prewhitened light curve. }
    \label{fig:FT260}
\end{figure}

\begin{figure}
    \includegraphics[width=0.5\textwidth]{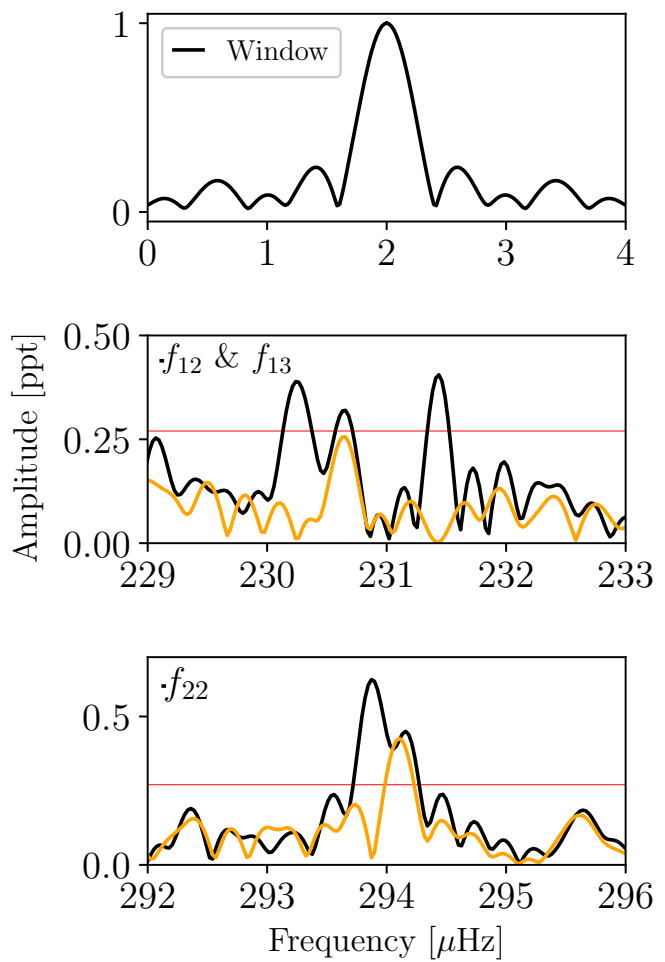} 
 \caption{Fourier transform of USC data of TIC\,260795163. The first raw shows the window function calculated from USC data. The second and third panels from the top display two different frequency regions in which the signals show strong residual after the extraction. These frequencies are given in Table \ref{260795163}.
 The horizontal red line indicates the 0.1\% FAP level. The orange line is the FT of the prewhitened light curve. 
 }
    \label{fig:FT_260_R}
\end{figure}

Concerning the g-mode region, we extracted 23 significant frequencies from both sector 1 and the extended mission (sector 27 and 28) dataset. 
The 13 frequencies that are found in both sectors (sector 1 and sector 27+28) are tabulated in Table \ref{260795163} and marked with $\dagger\dagger$. 
We found 7 frequencies in the extended mission dataset, which were not detected in sector 1. These frequencies are also tabulated in Table \ref{260795163} and marked with $\dagger$. 
As can be seen in the Fig. \ref{fig:FT260}, there are several frequencies above the threshold level in the extended mission (sector 27 and 28) which are not detected in sector 1 (see Table \ref{260795163} frequencies tagged with $\dagger$).

The amplitude spectrum of this object is dominated by a number of frequencies between 100 and 315 $\mu$Hz as it is shown in Fig.\ref{fig:FT260}. 
The top panel presents the amplitude spectrum from sector 1, while the bottom panel shows the amplitude spectrum from sectors 27 and 28.
In the lower frequency region, the four peaks (f$_{\rm 1}$, f$_{\rm 2}$, f$_{\rm 3}$ and f$_{\rm 8}$) were detected in sectors 27 and 28. 
The frequency f$_{\rm 5}$ is within 4.5 $\sigma$ level in 120-sec cadence data and in the USC data it is becoming a significant peak above 0.1\% FAP level.
However, there is a frequency, f$_{\rm 4}$, that is detected only in sector 1, with a S/N of 4.75, while it is not detected in sectors 27 and 28.
We recognized that there are residuals in the FT of the USC data after prewhitening. These residuals are shown in Fig. \ref{fig:FT_260_R}. 
The strong residuals at f$_{\rm 13}$ and f$_{\rm 23}$ could be either due to unresolved close-by frequencies or due to amplitude/frequency/phase variations over the length of the data. For these residuals, we did not prewhiten further. 
Overall, combining the nominal and extended mission dataset, we detect 23 g-modes spanning from 127 to 315 $\mu$Hz. 

{\bf TIC\,080290366} was found to be a pulsating star by \citet{koen2010}, who detected 5 oscillation frequencies ranging from 127\,$\mu$Hz to 233\,$\mu$Hz. 
TIC\,080290366 ($T_{mag} = 11.85$) was observed in SC mode during sector 2 (2018-Aug-22 to 2018-Sep-20) for 27.4\,d, with a frequency resolution of 0.63\,$\mu$Hz. Other parameters such as the length of the observations, contamination, number of data points and $0.1\%$ FAP confidence level are given in Table \ref{Table FT}. 
From sector 2, we detected 16 frequencies between 81 and 257 $\mu$Hz.

TIC\,080290366 was also observed during the extended mission in sector 29 (2020-Aug-26 to 2020-Sep-22).
The data length of sector 29 is 24.3 d, implying a lower frequency resolution of 0.72 $\mu$Hz. 
The FT average noise level is 0.067 ppt. The $0.1\%$ FAP confidence level is 0.316 ppt. 
We did not detect any high-frequency p-mode from the USC observations, while we detected 17 g-mode frequencies, most of them being present already in sector 2.
All the frequencies that were detected in both the nominal and extended mission are listed in Table \ref{080290366}. 
In total we have detected 18 frequencies spanning 
from 81.6\,$\mu$Hz ($\sim$12\,200\,s) to 257.4\,$\mu$Hz ($\sim$3\,900\,s) 
with amplitudes between 0.3 and 2.5 ppt.
Fifteen frequencies are detected in both datasets, 
the frequency near 141.7 $\mu$Hz was detected only in sector 2, while the three frequencies near
90.9, 114.7 and 168.0 $\mu$Hz were found only in sector 29.

\begin{figure}
    \includegraphics[width=0.5\textwidth]{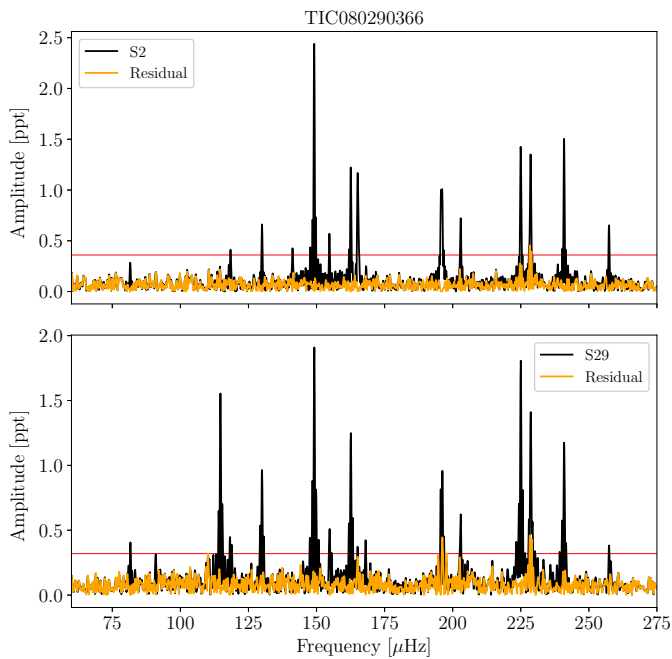} 
 \caption{{\sc Top:} Fourier transform of data taken in sector 2 of TIC\,080290366. 
 The horizontal red line indicates the 0.1\% FAP level. The orange line is the residual after extraction of the signals.
 {\sc Bottom:}  Fourier transform of data taken in sector 29 of TIC\,080290366. The horizontal red line indicates the 0.1\% FAP level. The orange line is the FT of the prewhitened light curve.}
    \label{fig:FT08}
\end{figure}

\begin{figure}
   \includegraphics[width=0.5\textwidth]{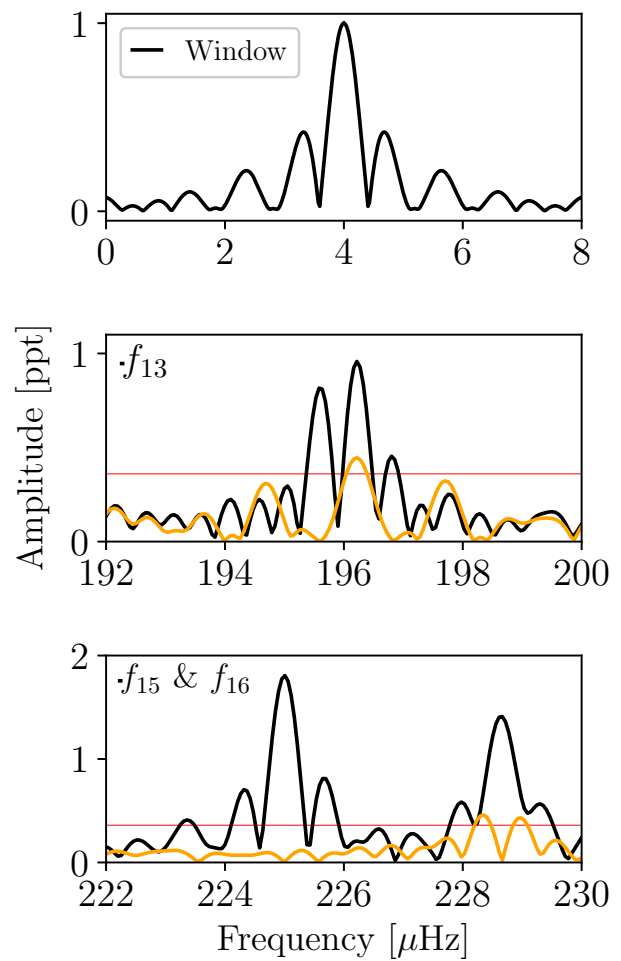}
 \caption{The Fourier Transform of sector 29 of TIC\,080290366, compared with the window function (top panel). The lower panels show an excess of power after prewhitening (orange line), see Section \ref{sect:sFT} for more details. The horizontal red line indicates the $0.1\%$ FAP level.}
 \label{fig:FT_08_R}
\end{figure}

The FTs of sector 2 (upper panel) and sector 29 (lower panel) are shown in Fig. \ref{fig:FT08}, where several frequencies clearly show differences.
In Fig. \ref{fig:FT_08_R}, we show the amplitude spectrum of sector 29 in more detail around the regions in which an excess of power is left after prewhitening, compared with the window function.

{\bf TIC\,020448010} ($T_{mag} = 12.18$) was observed in SC mode during sector 9 
for a period of 24.2\,d, which provides a frequency resolution of 0.717\,$\mu$Hz. The FT average noise level is 0.088 ppt. 
The rest of the parameters, including length of the observations, contamination, number of data points and $0.1\%$ FAP confidence level, are given in Table \ref{Table FT}. 
The star was discovered by \textit{TESS} as a g-mode sdB pulsator with 15 frequencies concentrated in a narrow region ranging from 68\,$\mu$Hz to 187\,$\mu$Hz abd with amplitudes between 0.36\,ppt and 1.26\,ppt. The extracted frequencies are listed in Table \ref{020448010} with their associated errors and S/N ratio. In Fig. \ref{fig:020448010}, we show all detected frequencies (light grey) and residuals (orange) after prewhitening.

{\bf TIC\,138707823}  ($T_{mag}$=12.7)  was  observed in SC mode during sector 2 between August 22 and September 20, 2018 for 27.4 days. 
From these SC observations of TIC\,138707823, we have extracted 7 periodicities from the light curve. 
All frequencies above $0.1\%$ FAP significance level of 0.563\,ppt are listed in Table \ref{138707823}. 
The frequencies are located in a narrow range 
from 87 to 237 $\mu$Hz. 

TIC\,138707823 was also observed during Sector 29 (2020-Aug-26 to 2020-Sep-22) with USC mode. The length of these observations (23.85 d) is almost 4 days shorter than the sector 2 dataset, resulting in somewhat worse frequency resolution (0.728 $\mu$Hz) than the one obtained for the SC dataset (0.634 $\mu$Hz).
The average noise level of the FT is 0.13 ppt. 
We extracted 3 significant frequencies above the threshold of 0.62 ppt. 
These 3 frequencies, which were also detected in sector 2, are marked with $\dagger\dagger$ in Table \ref{138707823}.
The four frequencies at 87, 132, 181 and 191 $\mu$Hz, which were not detected during the extended mission, are given without symbol in Table \ref{138707823}.

The amplitude spectrum of the sector 2 data of TIC\,138707823 is relatively poor in comparison with the four stars presented in this work, displaying only 7 frequencies above the threshold level. These seven frequencies can be seen in the upper panel of Fig. \ref{fig:138707823}. 
The bottom panel of the same figure shows only 3 significant frequencies from sector 29. 
Combining the results from sector 2 and 29, we detect 7 frequencies which are concentrated between 87\,$\mu$Hz and 237\,$\mu$Hz.  

\begin{figure}
    \includegraphics[clip,width=0.5\textwidth]{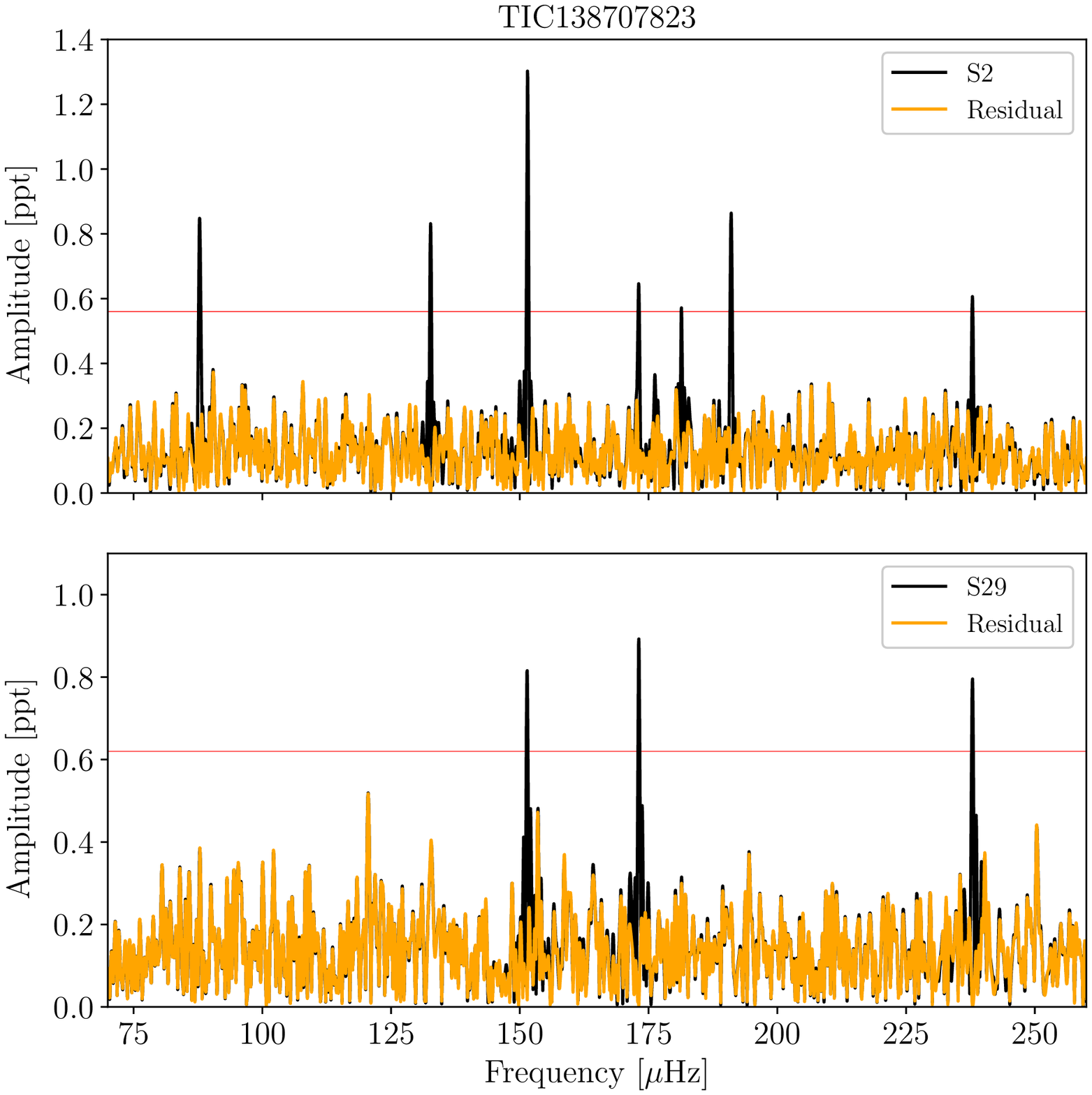}
 \caption{{\sc Top:} Fourier transform of sector 2 data of TIC\,138707823. The horizontal red line indicates the 0.1\% FAP level. The orange line is the residual after extraction of the signals. 
 {\sc Bottom:}  Fourier transform of sector 29 data of TIC\,138707823. The horizontal red line indicates the 0.1\% FAP level. The orange line is the FT of the prewhitened light curve.}
    \label{fig:FT138}
\end{figure}

{\bf TIC\,415339307} ($T_{mag} = 14.2$) was observed by \textit{TESS} in SC mode during sector 5 between 15 November 2018 and 11 December 2018, covering about 26 days.
The amplitude spectrum of this object, shown
in Fig. \ref{fig:415339307}, contains 9 frequencies above the $0.1\%$ FAP significance level of 1.27 ppt. The average noise level of the FT is 0.26 ppt. 
We note that there is a frequency at 184.122 $\mu$Hz (f$\rm_{3}$) just below the $0.1\%$ FAP confidence level, albeit at 4.5 $\sigma$, which we keep as a candidate frequency and discuss it's nature in section \ref{sect:sFT}. 

The photometric and FT parameters, including average noise level, contamination factor, number of data points and $0.1\%$ FAP confidence level, are given in Table \ref{Table FT}.
In Table \ref{415339307}, we list all (10) frequencies (periods) and their amplitudes with corresponding errors and we show all detected frequencies in Fig.\ref{fig:415339307}.

\subsection{Rotational multiplets}

The main goal of our analysis is to identify modes of detected pulsations in order to constrain theoretical models of pulsating sdB stars. 
For rotating stars, the existence of non-radial oscillations allows identification of the pulsation modes via rotational multiplets \citep{Aerts2010}.
The non-radial pulsations are described by three quantized numbers, $n, l$ and $m$, where $n$ is the number of radial nodes between center and surface, $l$ is the number of nodal lines on the surface, and $m$ is the azimuthal order, which denotes the number of nodal great circles that connect the pulsation poles of the star.
 
In rotating stars, the pulsation frequencies are split into 2$l +$1 azimuthal components due to rotation, revealing an equally spaced cluster of 2$l +$1 components. This 2$l +$1 configuration can be resolved with high-precision photometry if the star has no strong magnetic field and the rotational period is not longer than the duration of the observation.

Detection of rotational splitting is important as it is one of the two methods used to identify the pulsational modes of a star and at the same time it gives the information of the rotation period of the star. In particular for the g-mode pulsators, it unveils the rotation of the deep part of the radiative envelope close to the convective core.

In several sdBVs rotational multiplets have been detected \citep[][references therein]{Charpinet2018}. 
Typical rotation periods detected in sdB stars are of the order of 40 d \citep{barank22019}, unless the stars are in close binary systems.

Even though \textit{TESS} allows us to obtain uninterrupted time series, especially for stars observed in multiple sectors, it is not ideally suited for the detection of rotationally split multiplets in sdBVs.  
For the stars that have been observed in just 1 sector ($\sim$27 d), which translates in a frequency resolution of $\sim$ 0.6 $\mu$Hz, we are limited to the detection of rotational periods shorter than about 13 days. 
For each star, we have searched for a coherent frequency splitting, $\Delta \nu$ in the g-mode region and we did not find any consistent solution.
Therefore, we conclude that for neither of the sdBV stars analysed in this paper, it is possible to perform mode identification and determine rotational period based on rotational splitting. 
Given that 4 of the analyzed stars are not members of close binaries, we do not expect them to have short rotation periods such that they could be detected in a single sector \textit{TESS} data.
For TIC\,138707823 however, which is a short period binary system with an orbital period of about 4 d \citep{Edelmann2005}, we do not detect any significant signal that might be attributed to this orbital period.

\subsection{Frequency and amplitude variations}
\label{sect:sFT}

From the continuous light curves produced by the space missions such as \textit{Kepler} and K2 and now \textit{TESS}, it has been observed that oscillation frequencies in compact stars, including pulsating sdB stars, DBV and DOV pulsating white dwarfs, may not be stable \citep{silvottik22019,Zong_dbv_2016,corsico2020}.
It is known that the frequency and amplitude variations mostly occur due to beatings of unresolved peaks or unresolved multiplets.
Recently, the complex patterns that have been observed were interpreted as evidence of frequency/amplitude/phase modulations due to weak non-linear mode interactions, as discussed in \citet{ZCV2016}. Furthermore, the variability may occur due to the photon-count noise caused by contamination of the background light in the aperture.

The continuous photometric measurements of 5 stars allow us to construct sliding FTs (sFTs) to examine the temporal evolution of the detected frequencies over the course of the \textit{TESS} observations. 
Therefore, 
we have computed and examined the sFT of each target. 
Since three targets were observed in more than one sector, for these stars we selected the sectors in which we see the largest number of pulsation modes, i.e. sectors 27 and 28 for TIC\,260795163, sector 29 for TIC\,080290366, sector 2 for TIC\,138707823.

The sFTs are computed in a similar way as described in \citet{silvottik22019}: we run a 5-day sliding window with a step size of 0.2 days. The amplitudes are shown as color-scale in ppt units. We set a lower limit on the amplitudes by running the 3 times average noise level of 5-d chunk. 
Afterwards, we calculate the Fourier transform of each subset and trail them in time. 
  
\begin{figure}
    \includegraphics[clip,width=1.0\columnwidth]{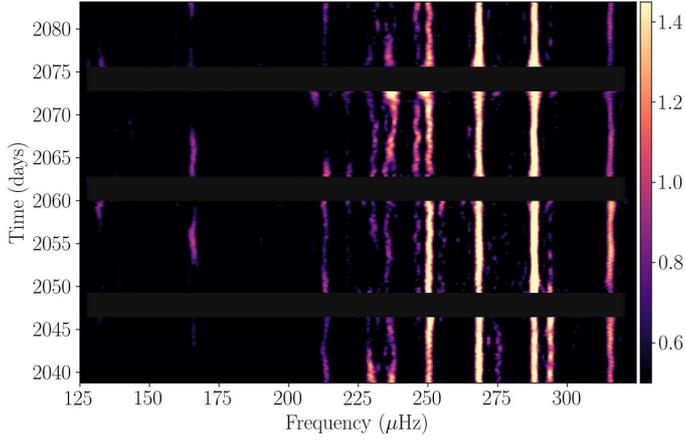}
 \caption{The  sliding  Fourier  Transform  (sFT)  of  TIC 260795163  from USC data.
 The color-scale amplitude is given in ppt units. See text for more details on sFT computations.
 }
    \label{fig:sft26}
\end{figure}

\begin{figure}
    \includegraphics[clip,width=1.0\columnwidth]{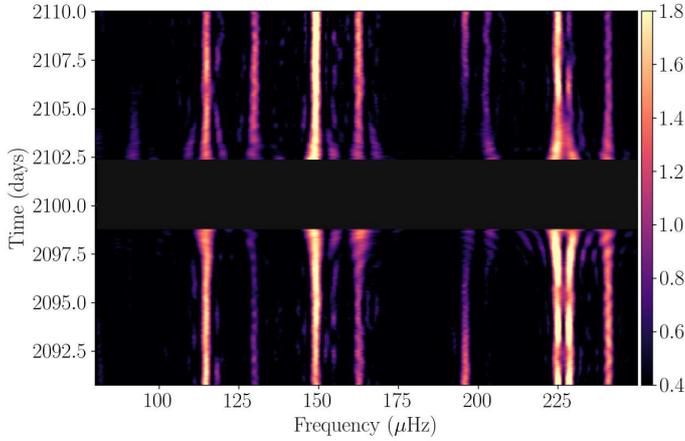}
 \caption{The sFT of TIC\,080290366. The sFT is calculated from USC data.}
    \label{fig:sft08}
\end{figure}

\begin{figure}
    \includegraphics[clip,width=1.0\columnwidth]{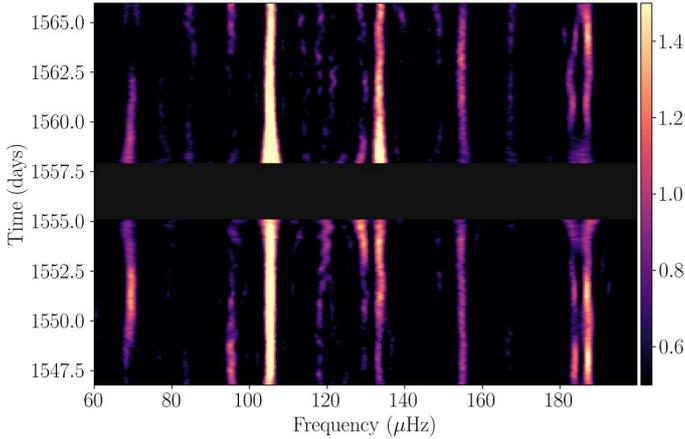} 
 \caption{The same figure as Fig. \ref{fig:sft08} but for TIC020448010. The sFT is calculated from SC data.}
    \label{fig:sft02}
\end{figure}

\begin{figure}
    \includegraphics[clip,width=1.0\columnwidth]{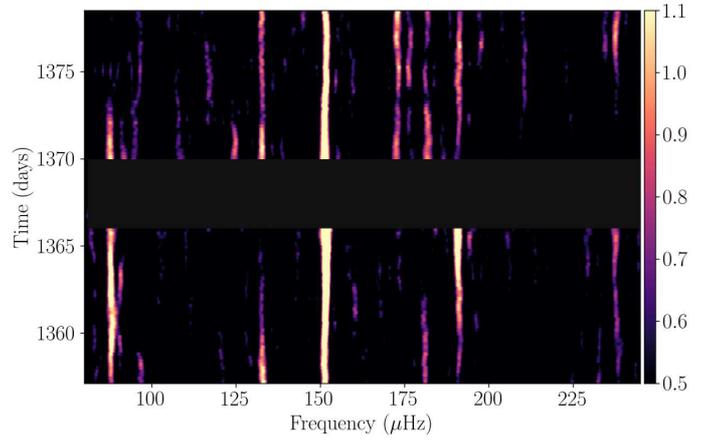} 
 \caption{The same figure as \ref{fig:sft08} but for TIC\,138707823. The sFT is calculated from SC data.}
    \label{fig:sft13}
\end{figure}

\begin{figure}
    \includegraphics[clip,width=1.0\columnwidth]{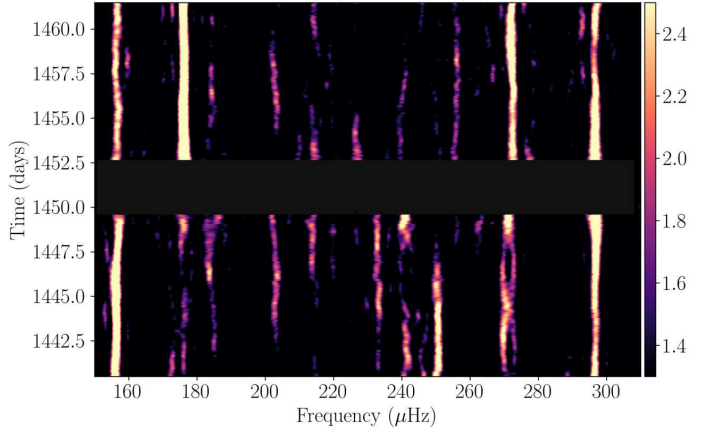}
 \caption{The same figure as \ref{fig:sft08} but for TIC\,415339307. The sFT is calculated from SC data.}
    \label{fig:sft41}
\end{figure}

In Fig. \ref{fig:sft26}, \ref{fig:sft08}, \ref{fig:sft02}, \ref{fig:sft13} and \ref{fig:sft41} we show the sFTs for the g-mode region of each star. 
In most of the cases, the high amplitude frequencies (S/N $\geq$ 10) are stable in both frequency and amplitude over the length of the data for all stars. 
However, some pulsational frequencies are not stable throughout the \textit{TESS} run and here we discuss each case. 

In the case of TIC\,260795163, the highest amplitude frequencies (at 250.447, 268.360, 288.186 and 315.537 $\mu$Hz) are stable in frequency, although the one at 250 $\mu$Hz shows a small wobble.
A few frequencies, mostly lower than 250 $\mu$Hz, are not stable, at least in amplitude. However, they have a low S/N and therefore are below the detection threshold throughout part of the run. 
In the case of the frequency at 293.996 $\mu$Hz (f$_{\rm 22}$), it is quite stable up to day $\sim$2050 and then it becomes weaker in amplitude. Indeed, we can see this effect in the bottom panel of Fig. \ref{fig:FT_260_R}, in which we show the strong residual after prewhitening, implying that either the amplitude, phase or frequency could be variable over the length of the data. 

In the case of TIC\,080290366 (see Fig. \ref{fig:sft08}), the amplitudes of the frequencies at 196.144 (f$_{\rm 14}$) and 228.658 $\mu$Hz (f$_{\rm 17}$) are variable and it is exactly these two frequencies that show residuals after prewhitening in Fig. \ref{fig:FT_08_R}. 

The highest amplitude frequency at 105.313 $\mu$Hz (f$_{\rm 5}$) of TIC\,020448010 is quite stable over the run, along with the frequency at 133.516 $\mu$Hz (f$_{\rm 10}$) (Fig. \ref{fig:sft02}). 
The low-amplitude frequencies are mostly unstable in amplitude, e.g. the frequency at 83.096 $\mu$Hz is absent between days 1551 and 1555.

In the case of TIC\,138707823 (Fig. \ref{fig:sft13}), we safely extracted 7 peaks from the SC observations and all the frequencies are visible in the sFT except for the frequency at 237.895 $\mu$Hz, whose S/N is low. 
The highest amplitude frequency at 151.497 $\mu$Hz is stable over the length of the data, while the rest of the frequencies (87.82, 132.67, 173.09, 181.38 and 191.02 $\mu$Hz) are not stable in amplitude, while they seem to be stable in frequency.

In the case of TIC\,415339307 (Fig. \ref{fig:sft41}), the frequency at 175.929 $\mu$Hz shows a strong variability in amplitude, being at very low S/N in the first half of the run 
while in the second half it has the highest amplitude of about 4 ppt. 
In Fig. \ref{fig:415339307}, this effect is seen as a significant residual at 5684 s. 
Frequencies at 203.126 and 214.359 $\mu$Hz are not stable in either frequency nor amplitude. 
In the case of the frequency at 184.122 $\mu$Hz (f$\rm_{3}$) (S/N = 4.45 in the entire FT), it is definitely present in the sFT, however it is not stable in either frequency nor amplitude during the first half of the run, while it appears more stable in frequency during the second half of the run. 
The frequency at 255.863 $\mu$Hz is stable up to almost 1448 d, while it is absent beyond 1448 d. Since this frequency is above the $0.1\%$ FAP level in the FT, we include it in our analysis as well.



We have not found strong evidence of rotational splitting in any of the targets analysed in this paper, and hence all of them must have rotation periods considerably longer than the length of their \textit{TESS} dataset. Consequently, any peak we have observed must be considered an unresolved multiplet, consisting of a summation of 3, for $l = 1$, or more sinusoids with independent phase and amplitude, and with each sinusoid having slightly different unresolved frequency. This produces beating effect on time scales longer than the analysed \textit{TESS} dataset, which then may appear as any form of frequency and/or amplitude variation. Beating of unresolved multiplets is therefore the default cause of any such variations  observed in datasets that are too short to reveal rotational splitting.

\subsection{Asymptotic g-mode period spacing}
\label{sect:APS}

\citet{fontaine2003} showed that the oscillation modes detected in long-period pulsating subdwarf B stars are associated with high-order g-modes. 
In the asymptotic limit, when {\it n}\,\texttt{>{}>}\,{\it l} the consecutive radial overtones of high-order g-modes are evenly spaced in period such that the consecutive g-modes follow the equation:
\begin{equation}
P_{l,n} = \frac{\Delta P_{0}}{\sqrt{l(l+1)}} n + \epsilon_{l}
\label{eq:power}
\end{equation}
where $\Delta P_0$ is just the asymptotic period spacing for g-modes, which is defined as $\Delta P_{0} \propto  [\int_{r_1}^{r_2} \frac{|N|}{r} dr]^{-1}$, $N$ being the Brunt–V\"ais\"al\"a frequency, the critical frequency of non-radial g-modes \citep{Tassoul1980} and
$\epsilon_{l}$ is a constant \citep{unno1979}.

The existence of a nearly constant period spacing of g-modes in the asymptotic regime means that we can search for the patterns of modes with a given harmonic degree in the observed period spectra of pulsating stars.
This method, so called the asymptotic period spacing method, has been used to identify the degree of pulsational modes in many g-mode pulsators such as $\gamma$-Dor stars, slowly pulsating B stars (SPBs) and white dwarfs \citep[][and references therein]{aerts2019}.  

However, only with the onset of space based light curves such as $Kepler$, this method became plausible for identifying the degrees of the pulsational modes in sdB stars \citep[e.g.]{reed2011}. 
However, with this method it is not possible to determine the absolute radial order {\it n} of a given mode from the equation \eqref{eq:power} without detail modelling. 
What we can do is choose a relative number for a radial order ($n_{l}$) corresponding to a modal degree $l$ such that all other consecutive radial orders for this modal degree will have an offset with respect to $n_{l}$.
The ratio between consecutive overtones is then derived from the equation \eqref{eq:power}, such that the ratio for dipole ({\it l}\,=\,1) and quadrupole ({\it l}\,=\,2) modes is $\sqrt{3}$. 
Based on the theoretical models by \citet{charpinet2000}, the period spacing of {\it l}\,=\,1 modes is around 250\,s for sdB pulsators.
The \textit{Kepler} and K2 observations of g-mode pulsating sdB stars find that the average period spacing for $l = 1$ modes is ranging from 227\,s to 276\,s \citep{Reed2018}. 
Recent results from the \textit{TESS} mission on five g-mode sdB pulsators find similar results, with the average period spacing ranging from 232\,s to 268\,s \citep{charpinet2019, saho2020, reed2020}. 
Here, we search for constant period spacing in our 5 target stars observed by \textit{TESS} using the Kolmogorov-Smirnov \citep[K-S;][]{kawaler1988} and the inverse variance \citep[I-V;][]{O'Donoghue1994} significance tests. 

In the K-S test, $Q$ is the quantity that defines the probability distribution of observed modes. 
If the data consist of non-random distribution in the period spectrum, then the distribution will have a peak at minimum value of $Q$. 
In the I-V test, on the other hand, a maximum of the inverse variance will indicate a consistent period spacing.  

\begin{figure}
    \centering 
    \includegraphics[clip,width=1.\linewidth]{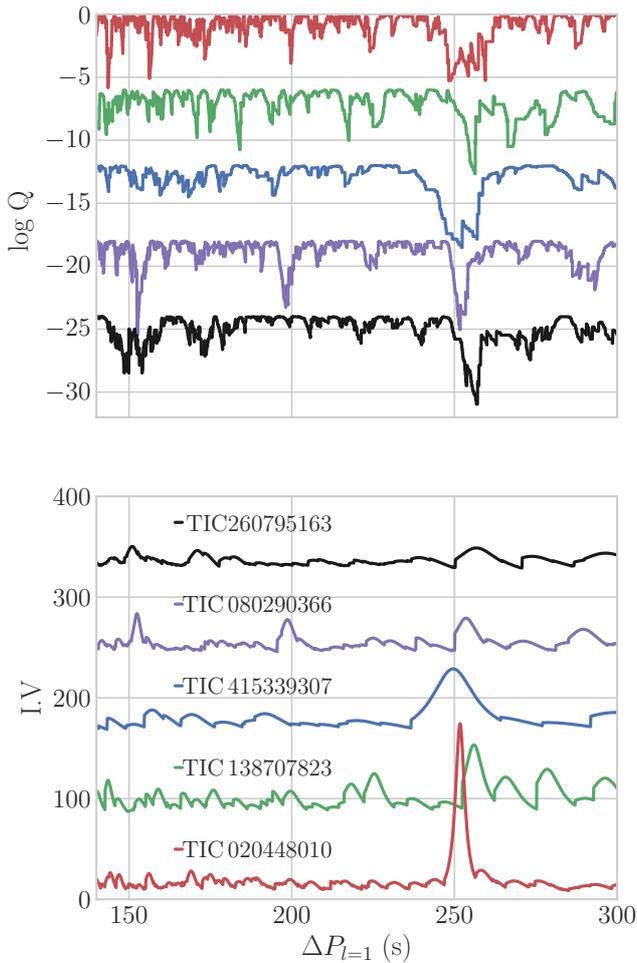}
      \caption{Kolmogorov-Smirnov (top panel) and Inverse-Variance (lower panel) tests to search for a constant period spacing in 5 stars. The offset on $log\,Q$ for the Kolmogorov-Smirnov is 6 and the offset on y-axis for the Inverse-Variance test is 80.
      All stars present a significant probability for a constant period spacing.}
    \label{fig:ks-iv}
\end{figure}

In Figure~\ref{fig:ks-iv}, we show the K-S (top panel) and I-V (bottom panel) results obtained for 5 stars. 
For both panels, we applied a vertical, arbitrary offset, for a visualization purpose. The same color coding is applied for both panels.
For all targets, the statistical tests display a clear indication of the mean period spacing at around 250 s. 
For the case of TIC\,260795163 and TIC\,080290366, the I-V test is not as conclusive as for the other 3 stars. 
However, the K-S test does show an indication for a possible mean period spacing of $l=1$ at 250 s.
Moreover, TIC\,260795163 and TIC\,080290366 show peaks at around 150 s referring to the possible period spacing of quadrupole modes.

Based on the potential period spacings obtained from the K-S and I-V tests, we search for the sequences of dipole and quadrupole modes in the entire FT of each star. 
First, we search for the $l = 1$ and $l = 2$ sequences examining consecutive modes in period domain of the FT of each star. 
Given that the higher degree modes are more sensitive to geometrical suppression due to mode cancellation effect \citep{Aerts2010}, we assume that the highest amplitude frequencies are corresponding to low-degree modes ($l=1$ and $l=2$). 
This assumption is valid only if all the modes have the same intrinsic amplitude. 
For sdB pulsators, the majority of detected frequencies have been identified with low-degree modes ($l\leq$ 2).
However, a few exceptional examples have been reported. 
For instance, in two sdB pulsators observed during the nominal mission of \textit{Kepler}, the high-degree g-modes were assigned up to $l = 6$ \citep{kern2018} or $l = 8$ \citep{telting2014} using the method of rotational multiplets.
\citet{silvottik22019} also reported several high-degree modes up to $l = 12$ 
using solely asymptotic period spacing in the brightest ($V=10.2$) sdB pulsator HD\,4539 (EPIC\,220641886) observed during the K2 mission.
We proceed in the following way: we assign arbitrary radial orders ($n$) for each identified $l$ degree and calculate the mean period spacing using a linear regression fit. 
In this way we have calculated the mean period spacing ($\Delta P$) for all stars and found that the mean period spacing for the $l = 1$ mode is ranging from 251\,s and 255\,s.
In order to assess the errors of the mean period spacing obtained in our analysis, we perform a bootstrap resampling analysis as described by \citet{efron1979, Simpson1986}. 
We used this method because many possible modes are not detected in the amplitude spectra. 
Furthermore, in some cases the individual pulsational period has no unique modal degree solution, and 
in some specific cases the modes could be altered due to mode trapping.
 
In order to make a realistic error assignment, we simulated $10^{4}$ datasets from the determined $l$ modes. 
For each target, we created sets of randomly chosen observed periods that are already identified as $l = 1$ or $l = 2$ modes, in order to obtain the mean period spacing from each different sub-sequences. 
The same data point can occur multiple times and ordering is not important, such that for $N$ data points the total number of possible different bootstrap samples is 
$\frac{(2N - 1)!}{N!(N - 1)!}$ \citep{rene2010}. 
For instance, for a given dipole sequence, which consists of 10 modes, the total number of potential different bootstrapping includes about $10^{5}$ subsets. 
For 5 stars analyzed in this paper, we detected the dipole sequences that range from 7 to 13 modes depending on a star. 
For that reason, we restricted ourselves with $10^{4}$ subsets. 
For each such subset which includes a series of dipole or quadrupole modes, we derive the mean period spacing with the linear regression fit. 
The most probable solution is obtained as a mean period spacing (which corresponds to the 50 percentile of the $\Delta P_{l=1}$ distribution).
The errors are then estimated as 1 $\sigma$ and given in Table \ref{Table Seismic}.

The right panels of Figure~\ref{fig:260795163}, \ref{fig:080290366}, \ref{fig:020448010}, \ref{fig:138707823} and \ref{fig:415339307} show the residuals between the observed periods and the periods derived from the mean period spacing for the $\Delta P_{l=1}$ in where we can see the deviation of the modes. 
The scatter of the residuals for all stars is up to 50 s and for the stars that we have more $l = 1$ modes detected, we notice the oscillatory pattern which is a characteristic feature that was found in several V1093\,Hya type sdB pulsators \citep[e.g.]{telting2012,baran2012a}. 
Detecting all modes of $l = 1$ sequence with an expected period spacing of 250\,s \citep{charpinet2002} is unlikely as the sdB stars are chemically stratified, which causes the observed modes to be scattered around this value.
Such small deviations from the mean period spacing are to be expected in those stars where diffusion processes have had enough time to smooth out the H-He transition zone \citep{MillerBertolami2012}. 
On the other hand, the efficiency of trapping diminishes with increasing the radial order as discussed in \citet{2014ASPC..481..179C}. This is because the local wavelength of the modes decreases with increasing the radial order and therefore, the higher order g-modes become less affected by the H-He transition zone. In this way the higher order g-modes may present the nearly constant period spacing even without having the H-He transition zone smoothed.  

\subsubsection{TIC\,260795163}

The K-S test for TIC\,260795163 shows an indication for both dipole and quadrupole modes, see  the upper panel of Fig. \ref{fig:ks-iv}. 
Assuming that the highest amplitude frequency at 288.186 $\mu$Hz (f$_{\rm 21}$ in Table \ref{260795163}) is an $l = 1$ mode, then the sequence of $l = 1$ mode is fulfilled with the following frequencies: f$_{\rm 10}$, f$_{\rm 14}$, f$_{\rm 17}$ and f$_{\rm 19}$. 
Beyond 6000 s, 6 additional frequencies (f$_{\rm 1}$ to f$_{\rm 6}$)  were found to fit the dipole sequence.
The following six frequencies: f$_{\rm 8}$, f$_{\rm 9}$, f$_{\rm 11}$, f$_{\rm 15}$, f$_{\rm 16}$ and f$_{\rm 20}$ showed consistent period spacing for quadrupole modes with a unique solution. 
For 4 frequencies we could not find a unique solution, namely f$_{\rm 4}$, f$_{\rm 6}$, f$_{\rm 10}$ and f$_{\rm 21}$, and they could be interpreted either as $l = 1$ or $l = 2$ modes.  
  
\begin{figure*}
    \centering    
    \includegraphics[clip,width=1.\linewidth]{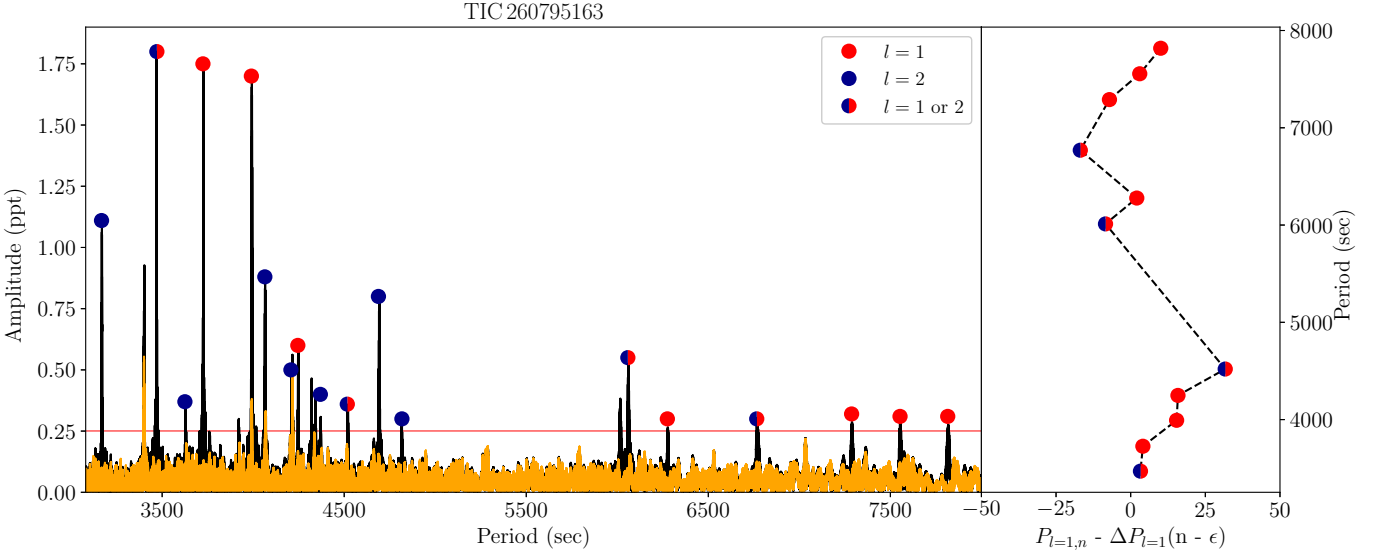} 
 \caption{The FT of TIC\,260795163 (presented here are the sectors 1, 28 and 29 together). 
 The residuals after prewhitening are shown with orange color. The red dots are showing the dipole modes and the blue ones are displaying the quadrupole modes. If there is no unique identification, the period is shown with both colors. The horizontal red line correspond to $0.1\%$ FAP confidence level.
 The right panel shows the residuals between the observed and the fitted periods.
 See text for more details on mode identification and the mean period spacing computations.
 }
   \label{fig:260795163}
\end{figure*}

Out of the total of 23 periodicities detected in 
\textit{TESS} data of TIC\,260795163, there were 5 frequencies that did not fit neither $l = 1$ nor $l = 2$ sequences. 
These periodicities could be either higher-degree modes or  $l = 1$ and/or $l = 2$ modes which are severely affected by mode trapping and do not fit the $l = 1$ and $l = 2$ patterns. 
However, without detecting rotational multiplets, it is impossible to test any of the two possibilities. 
The amplitude spectra of TIC\,260795163 with our mode identification are presented in Fig. \ref{fig:260795163}. 
Based on this mode identification, we calculate the mean period spacing of dipole modes, $\Delta P_{l=1}$ = $254.83^{+2.14}_{-2.28}$ s and quadrupole modes, $\Delta P_{l=2}$ = $150.51^{+3.27}_{-2.28}$ s. 

\begin{table}
\setlength{\tabcolsep}{2pt}
\renewcommand{\arraystretch}{1.1}
\centering
\caption{Frequency solution from the \emph{TESS} light curve of TIC\,260795163 including frequencies, periods, and 
amplitudes (and their uncertainties) and the signal-to-noise ratio. Errors are given in parenthesis to 2 significant digits. Identified modal degree and relative radial orders are listed column 5 and 6, respectively.
The frequencies detected in both sector 1 and the extended mission of sector 27 and 28 are tagged with $\dagger\dagger$.
The frequencies detected only in sector 27 and 28  are tagged with $\dagger$. 
The frequencies that are detected only in sector 1 are given without tag. 
See text for more details on mode identification.}
\begin{tabular}{cccccccr}
\hline
\noalign{\smallskip}
ID & Frequency & Period & Amplitude & S/N & $l$ & $n$  \\
   & $\mu$Hz   &  [sec] &  [ppt]    &     &     &       \\
\noalign{\smallskip}
\hline
\noalign{\smallskip}
f$_{\rm 1^{\dagger}}$ &  127.915 (20) &  7817.63 (1.27)	 &  0.269(44) & 4.84    &  1   &  30  &      \\
f$_{\rm 2^{\dagger}}$ &  132.358 (14) &  7555.25 (81)    &  0.394(44) & 7.09    &  1   & 29   &      \\
f$_{\rm 3^{\dagger}}$ &  137.178 (16) &  7289.77 (85)    &  0.349(44) & 6.28    &   1  &  28  &      \\
f$_{\rm 4}$ &    147.752 (36)  &   6768.08 (1.66)  &   0.475 (76) &  4.75  &   1/2  &  26/44   &      \\  
f$_{\rm 5}$ &   159.298 (10)  &   6277.541  (35)  &  0.26 (43)  &    4.74   &   1   &    24   \\ 
f$_{\rm 6^{\dagger\dagger}}$ &  165.001 (10) &  6060.55 (37)    &  0.554(44) & 9.98    & 1/2  & 23/39      \\
f$_{\rm 7^{\dagger}}$ &  166.194 (14) &  6017.04 (53)    &  0.394(45) & 7.09    &     &    &      \\
f$_{\rm 8}$ &   207.634 (10)  &   4816.172  (20)  &  0.28 (43)  &   5.17   &    2  &    31   \\ 
f$_{\rm 9^{\dagger\dagger}}$ &  213.093 (08) &  4692.77 (18)    &  0.657(44) & 11.83   &  2   & 30       \\
f$_{\rm 10^{\dagger}}$ &  221.251 (14) &  4519.75 (29)    &  0.392(44) & 7.06    &  1/2   &  17/29 &      \\
f$_{\rm 11^{\dagger}}$ &  228.75	(12)  &  4371.58 (23)    &  0.465(44) & 8.36    &  2   &  28  &      \\
f$_{\rm 12^{\dagger\dagger}}$ &  230.251 (15) &  4343.08 (29)    &  0.356(44) & 6.41    &     &         \\
f$_{\rm 13^{\dagger\dagger}}$ &  231.428 (13) &  4320.98 (24)    &  0.428(44) & 7.71    &     &        \\
f$_{\rm 14^{\dagger\dagger}}$ &  235.444 (08) &  4247.28 (14)    &  0.683(44) & 12.28   &   1  &  16       \\
f$_{\rm 15^{\dagger\dagger}}$  &  237.532 (10) &  4209.94 (18)    &  0.536(44) & 9.64    &  2   & 27   &      \\
f$_{\rm 16^{\dagger\dagger}}$ &  245.960 (08) &  4065.69 (14)    &  0.654(44) & 11.77   &  2   & 26      \\
f$_{\rm 17^{\dagger\dagger}}$ &  250.447 (04) &  3992.84 (06)    &  1.349(44) & 24.27   &  1   & 15       \\
f$_{\rm 18^{\dagger}}$ &  254.972 (18) &  3921.98 (28)    &  0.308(44) & 5.54    &     &         \\ 
f$_{\rm 19^{\dagger\dagger}}$ &  268.360 (03) &  3726.32 (04)    &  1.624(44) & 29.20   &  1   & 14         \\
f$_{\rm 20^{\dagger\dagger}}$ &  275.588 (16) &  3628.59 (22)    &  0.335(44) & 6.03    &   2  &  23        \\
f$_{\rm 21^{\dagger\dagger}}$ &  288.186 (03) &  3469.97 (03)    &  1.716(44) & 30.87   &  1/2   &  13/22       \\
f$_{\rm 22^{\dagger\dagger}}$  &   293.996 (12)  &   3401.44 (11)  &   1.469 (76) &  14.70   &       &              \\
f$_{\rm 23^{\dagger\dagger}}$ &  315.537 (05) &  3169.19 (05)    &  1.043(44) & 18.77   &  2   &  20       \\
\noalign{\smallskip}
\hline
\label{260795163}
\end{tabular}
\end{table}

\subsubsection{TIC\,080290366}

\begin{figure*}
    \centering    
    \includegraphics[clip,width=1.\linewidth]{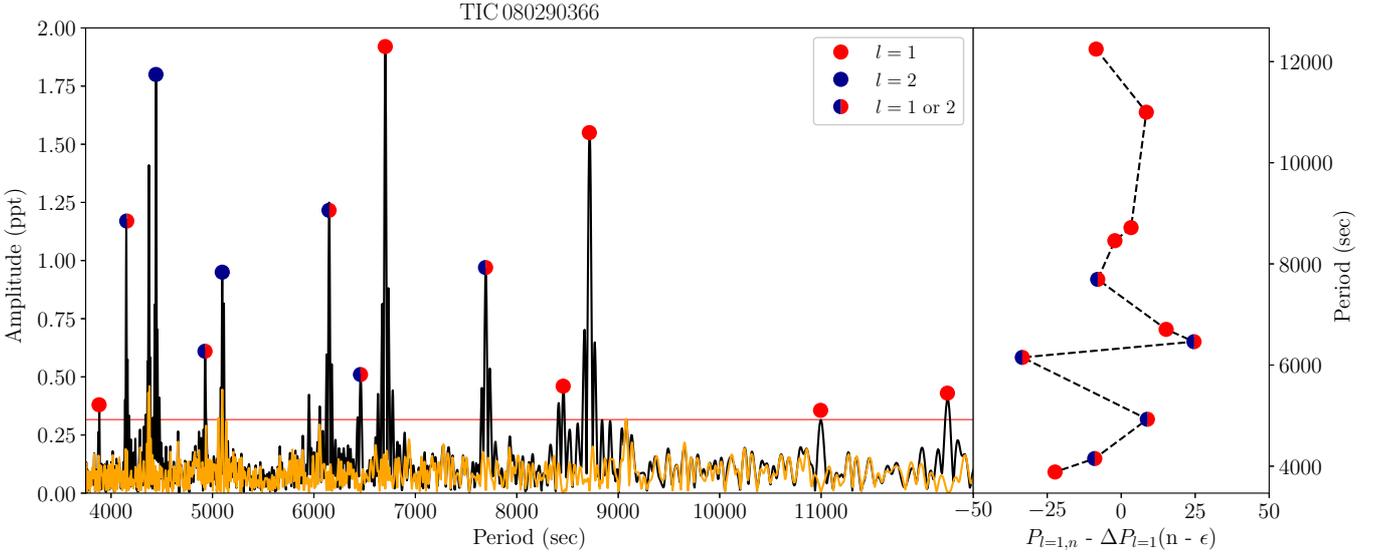} 
 \caption{Same as Fig. \ref{fig:260795163} but for TIC\,080290366 (presented here is sector 29). }
   \label{fig:080290366}
\end{figure*}

Based on the results from the K-S and I-V tests in which there is a peak at around 255\,s, indicating a possible $\Delta P_{l=1}$ and also a peak at 155\,s indicating a possible $\Delta P_{l=2}$ (see upper and lower panel of Fig. \ref{fig:ks-iv}) we searched for dipole and quadrupole sequences in the FT of TIC\,080290366.  
Starting from the highest amplitude frequency at 149.133 $\mu$Hz (f$_{\rm 7}$ in Table \ref{080290366}) as an $l = 1$ mode, we find two frequencies that fit the dipole mode sequence, f$_{\rm 8}$ and f$_{\rm 9}$. Beyond 7\,500\,s, there are 5 frequencies which are also following dipole mode sequences: f$_{\rm 1}$, f$_{\rm 2}$, f$_{\rm 3}$, f$_{\rm 4}$ and f$_{\rm 5}$. 
We found only 3 frequencies that uniquely fit the $l = 2$ sequence. 
For 5 frequencies (f$_{\rm 5}$, f$_{\rm 8}$, f$_{\rm 9}$, f$_{\rm 14}$ and f$_{\rm 17}$) the solution is degenerate as the frequencies fit both dipole and quadrupole mode sequences. 
Out of 18 detected frequencies there are 4 frequencies that did not fit dipole nor quadrupole mode sequences. These frequencies could be high-order degree modes or trapped modes.
The amplitude spectra of TIC\,080290366 with our mode identification are presented in Fig. \ref{fig:080290366}. 
Based on the above explained mode identification, we calculated both $\Delta P_{l=1}$ =  $253.32^{+0.78}_{-0.84}$ s and $\Delta P_{l=2}$ = $154.16^{+2.47}_{-7.97}$ s.
The final seismic result for TIC\,080290366 is given in Table \ref{Table Seismic}.

\begin{table}
\setlength{\tabcolsep}{1.8pt}
\renewcommand{\arraystretch}{1.1}
\centering
\caption{Frequency solution from the \emph{TESS} light curve of TIC\,080290366 including frequencies, periods, and 
amplitudes (and their uncertainties) and the signal-to-noise ratio. Errors are given in parenthesis to 2 significant digits. Identified modal degree and relative radial orders are listed in column 5 and 6, respectively.
The frequencies detected in both sector (2 and 29) are tagged with $\dagger\dagger$.
The frequencies detected only in sector 29 are tagged with $\dagger$. 
The frequencies that are detected only in sector 2 are given without tag. 
See text for more details on mode identification.
}
\begin{tabular}{cccccccr}
\hline
\noalign{\smallskip}
ID & Frequency & Period & Amplitude & S/N & $l$ & $n$  \\
   & $\mu$Hz   &  [sec] &  [ppt]    &     &     &       \\
\noalign{\smallskip}
\hline
\noalign{\smallskip}
f$_{\rm 1^{\dagger\dagger}}$ & 81.649  (31) &	12247.525 (4.62)  & 0.412 (53)	&   6.87    & 1 & 47 &   \\
f$_{\rm 2^{\dagger}}$  &       90.912 (41)	&	10999.552 (4.86)   & 0.313 (53) & 5.3 & 1 & 42  \\ 
f$_{\rm 3^{\dagger}}$ & 114.712	(08) &	8717.476  (63)	  & 1.560	(53)	&   26.00   & 1 & 33 &   \\
f$_{\rm 4^{\dagger\dagger}}$ & 118.217	(28) &	8459.017  (2.06)  &	0.452	(53)	&   7.54    & 1 & 32 &   \\
f$_{\rm 5^{\dagger\dagger}}$  & 129.967	(13) &	7694.241  (78) 	  & 0.982	(53)	&   16.37   & 1/2 & 29/48 &   \\
f$_{\rm 6}$ &   141.123 (30)  &   7086.02  (1.45)  &  0.390 (52)  &    6.11   &   2   &  44     \\
f$_{\rm 7^{\dagger\dagger}}$  &   149.133 (5)   &   6705.43  (22)  &  2.452 (52)  &    37.02  &   1   &  25       \\
f$_{\rm 8^{\dagger\dagger}}$  & 154.754	(24) &	6461.854  (1.02)  & 0.532	(53)	&   8.87    & 1/2 & 24/40   \\
f$_{\rm 9^{\dagger\dagger}}$  & 162.579 (10) &	6150.830  (38)    & 1.266 (53)	&   21.11   & 1/2 & 23/38   \\
f$_{\rm 10^{\dagger\dagger}}$ &   165.129 (10)  &   6055.88  (62)  &  1.142 (52)  &    17.41  &      &         \\
f$_{\rm 11^{\dagger}}$ & 168.027 (02) &	5951.40  (43)	  & 0.371	(53)	&   5.55   &   &    \\
f$_{\rm 12^{\dagger\dagger}}$  &  195.836 (34)  &   5106.29  (88)  &  1.51 (51)  &   22.98  &      &       \\
f$_{\rm 13^{\dagger\dagger}}$  &  196.144 (11)  &   5098.29  (30)  &  1.03 (52)  &   15.63  &   2   &  31     \\
f$_{\rm 14^{\dagger\dagger}}$ &  202.916 (16)  &   4928.14  (39)  &  0.737 (52)  &    11.11  &   1/2   &  18/30    \\
f$_{\rm 15^{\dagger\dagger}}$ & 224.994 (07) &	4444.544  (14)    & 1.784	(53)	&   29.73   & 2 & 27 &   \\
f$_{\rm 16^{\dagger\dagger}}$  & 228.658	(09) &	4373.344  (18)    & 1.392	(54)	&   23.20   &  &  &   \\
f$_{\rm 17^{\dagger\dagger}}$ &  240.884 (8)   &   4151.37  (13)  &  1.518 (52)  &    22.93  &   1/2 &  15/25 &        \\
f$_{\rm 18^{\dagger\dagger}}$ &  257.402 (18)  &   3884.97  (27)  &  0.635 (51)  &    9.76   &   1   &  14    &        \\
\noalign{\smallskip}
\hline
\label{080290366}
\end{tabular}
\end{table}

\subsubsection{TIC\,020448010}

The K-S and IV tests display a clear indication of a period spacing at 253\,s as it is shown in Fig. \ref{fig:ks-iv}. 
Starting with the highest amplitude peak at 9\,495.5\, s (f$_{\rm 5}$ in Table \ref{020448010}), we search for a dipole sequence in the FT. Out of 15 detected frequencies, 12 frequencies uniquely fit the $l = 1$ sequence, while only 2 frequencies can be evaluated as $l = 2$ mode as the difference is corresponding to 2 x $\Delta P_{l=2}$. 
One periodicity (at 8778.2 s) however, does not have a unique solution, it could be either $l = 1$ or $l = 2$ mode.

\begin{figure*}
    \centering    
    \includegraphics[clip,width=1.\linewidth]{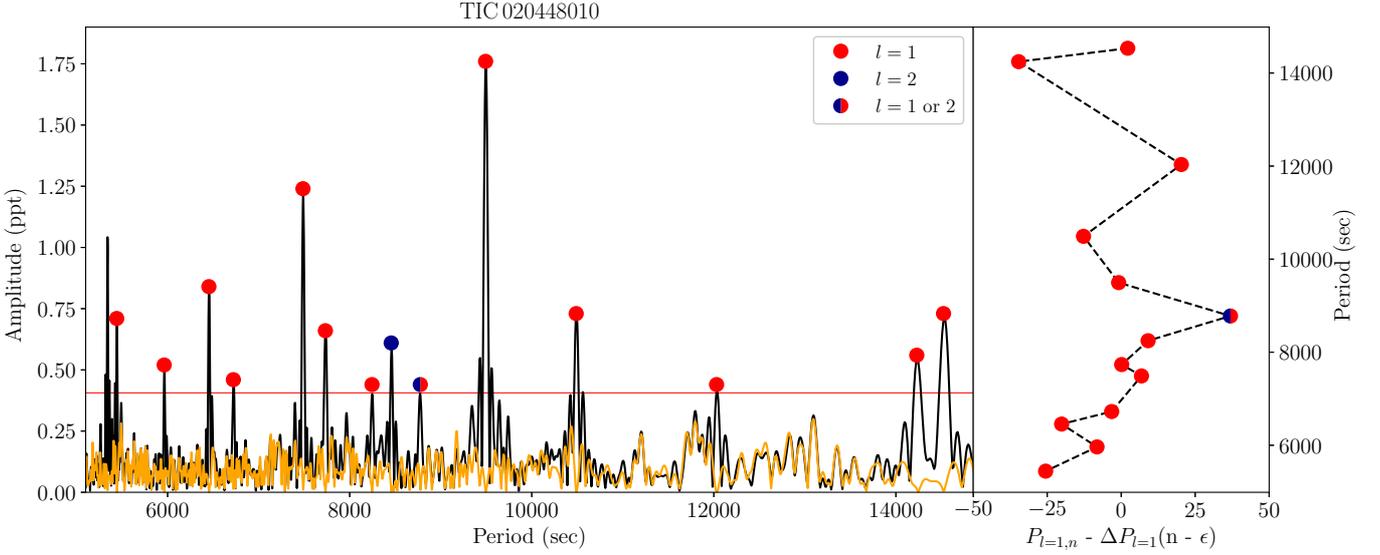} 
 \caption{Same as Fig. \ref{fig:260795163} but for TIC\,020448010 (presented here is sector 9).}
   \label{fig:020448010}
\end{figure*}

The amplitude spectra of TIC\,020448010 together with the mode identification are presented in Fig. \ref{fig:020448010}.  
Based on our mode identification we calculate the average period spacing for dipole modes, $\Delta P_{l=1}$ = $251.70^{+0.87}_{-0.96}$ s. 
In Table \ref{Table Seismic}, we also gave $\Delta P_{l=2}$ using the ratio between consecutive overtones derived from the equation \eqref{eq:power}.

\begin{table}
\setlength{\tabcolsep}{2pt}
\renewcommand{\arraystretch}{1.1}
\centering
\caption{Frequency solution from the \emph{TESS} light curve of TIC\,020448010 including frequencies, periods, and 
amplitudes (and their uncertainties) and the signal-to-noise ratio. Identified modal degree and relative radial orders are listed column 5 and 6, respectively.}
\begin{tabular}{cccccccr}
\hline
\noalign{\smallskip}
ID & Frequency & Period & Amplitude & S/N & $l$ & $n$  \\
   & $\mu$Hz   &  [sec] &  [ppt]    &     &     &       \\
\noalign{\smallskip}
\hline
\noalign{\smallskip}                                                  
 f$_{\rm 1}$ &   68.845(24) &  14525.3(5.1) &   0.731(7) &   8.5 &  1    &  56    &   \\ 
 f$_{\rm 2}$ &   70.203(32) &  14244.3(6.5) &   0.561(7) &   6.5 &  1    &  55     &   \\ 
 f$_{\rm 3}$ &   83.096(46) &  12034.3(6.7) &   0.393(7) &   4.4 &  1    &  46    &   \\ 
 f$_{\rm 4}$ &   95.327(25) &  10490.2(2.7) &   0.742(7) &   8.2 &  1    &  40    &   \\ 
 f$_{\rm 5}$ &  105.313(10) &   9495.5(9)   &   1.752(7) &  20.4 &  1    &  36    &   \\ 
 f$_{\rm 6}$ &  113.918(49) &   8778.2(3.7) &   0.358(7) &   4.2 &  1/2  &  33/49    &   \\ 
 f$_{\rm 7}$ &  118.227(28) &   8458.3(2.0) &   0.615(7) &   7.3 &  2  &  48    &   \\ 
 f$_{\rm 8}$ &  121.244(39) &   8247.8(2.7) &   0.442(7) &   5.2 &  1  &  31    &   \\ 
 f$_{\rm 9}$ &  129.282(27) &   7735.0(1.6) &   0.677(7) &   7.7 &  1  &  29    &   \\ 
 f$_{\rm 10}$ & 133.516(14) &   7489.7(8)   &   1.260(7) &  14.6 &  1  &  28    &   \\ 
 f$_{\rm 11}$ & 148.722(38) &   6724.0(1.7) &   0.449(7) &   5.3 &  1  &  25    &   \\ 
 f$_{\rm 12}$ & 154.898(21) &   6455.9(9)   &   0.865(7) &   9.8 &  1  &  24    &   \\ 
 f$_{\rm 13}$ & 167.668(34) &   5964.2(1.2) &   0.508(7) &   6.0 &  1  &  22    &   \\ 
 f$_{\rm 14}$ & 183.706(22) &   5443.5(6)   &   0.798(7) &   9.4 &  1  &  20    &   \\ 
 f$_{\rm 15}$ & 187.244(16) &   5340.64(46) &   1.122(7) &  12.6 &    &    &   \\ 
\noalign{\smallskip}
\hline
\label{020448010}
\end{tabular}
\end{table}

\subsubsection{TIC\,138707823}

\begin{figure*}
    \centering    
    \includegraphics[clip,width=1.\linewidth]{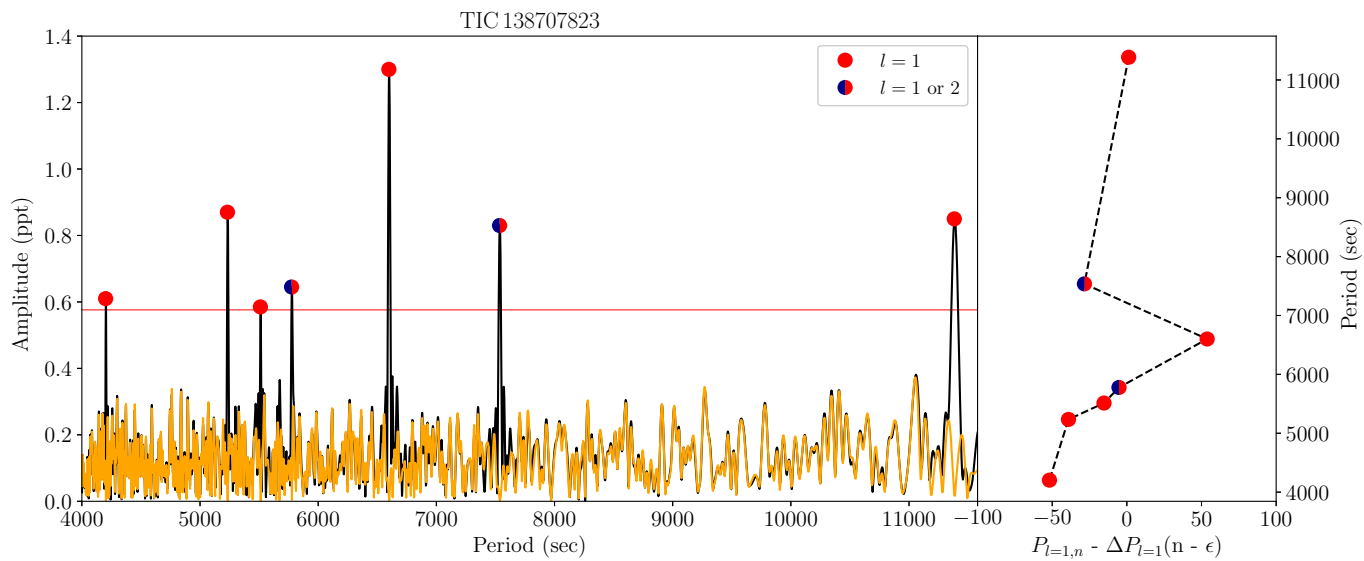} 
 \caption{Same as Fig. \ref{fig:260795163} but for TIC\,138707823 (presented here is sector 2).}
   \label{fig:138707823}
\end{figure*}

The K-S and IV tests show a clear indication of a possible $\Delta P_{l=1}$ at 255\,s as can be seen in Fig. \ref{fig:ks-iv}. 
Assuming the highest peak at 6\,600.78\,s (f$_{\rm 3}$) is an $l = 1$ mode, we find that the 
dipole sequence can be completed with f$_{\rm 4}$, f$_{\rm 5}$ and f$_{\rm 6}$. 
In the longer period region beyond 6\,600\,s, there are two more frequencies 
with the large gap in between that fit the $l = 1$ sequence. We interpreted these two frequencies as $l = 1$ modes and added them to the fitting procedure in order to find the mean period spacing of $\Delta P_{l=1}$. 
The same case is valid for the shorter period region and between f$_{\rm 6}$ and f$_{\rm 7}$ there is a big gap, of about 1\,000\,s (which is 4 x $\Delta P_{l=1}$). 
We noted that the periodicity (at 6600.78 s) is deviated from the mean period spacing significantly. This mode can be interpreted as a candidate of trapped mode, however, without knowing where the quadrupole modes, it is not allowed to be conclusive.
The two frequencies f$_{\rm 2}$ and f$_{\rm 4}$ could  be also interpreted as $l = 2$ modes since three times $\Delta P_{l=1}$ equals to five times $\Delta P_{l=2}$. 
If we exclude these two degenerate modes and calculate the average period spacing of $l = 1$, we find $\Delta P_{l=1}$ = 256.38 $\pm$ 1.43\,s. 
If we include these two modes and calculate the average period spacing, we found $\Delta P_{l=1}$ = $ 256.09^{+6.97}_{-1.21}$ s. 
We list all modes identified in Table \ref{138707823} and show them in Fig. \ref{fig:138707823}.

The final seismic result for TIC\,138707823 and the rest of the stars analyzed in this paper are given in Table \ref{Table Seismic}.

\begin{table}
\setlength{\tabcolsep}{2pt}
\renewcommand{\arraystretch}{1.1}
\centering
\caption{Frequency solution from the \emph{TESS} light curve of TIC\,138707823 including frequencies, periods, and 
amplitudes (and their uncertainties) and the signal-to-noise ratio. Errors are given in parenthesis to 2 significant digits. Identified modal degree and relative radial orders are listed column 5 and 6, respectively.
The frequencies detected in both sector (2 and 29) tagged with $\dagger\dagger$.
The frequencies detected only in sector 29 are tagged with $\dagger$. 
The frequencies that are detected only in sector 2 data given without tag. 
See text for more details on mode identification.}
\begin{tabular}{cccccccr}
\hline
\noalign{\smallskip}
ID & Frequency & Period & Amplitude & S/N & $l$ & $n$  \\
   & $\mu$Hz   &  [sec] &  [ppt]    &     &     &       \\
\noalign{\smallskip}
\hline
\noalign{\smallskip}  
f$_{\rm 1}$ & 87.829  (26) &  11385.70 (3.33)  &  0.858 (97)  &   6.98   &  1    &  44    \\
f$_{\rm 2}$ & 132.679 (27) &  7537.01  (1.51)  &  0.833 (97)  &   6.77   &  1/2  &  29/45    \\
f$_{\rm 3^{\dagger\dagger}}$ & 151.497 (17) &  6600.78  (74)    &  1.297 (97)  &   10.55  &  1    &  25    \\
f$_{\rm 4^{\dagger\dagger}}$ &173.099 (29)	&  5777.02	(97)    &	0.890	(10)	&  6.84  & 1/2 & 22/33 \\
f$_{\rm 5}$ & 181.380 (39) &  5513.27  (1.17)  &  0.576 (98)  &   4.68   &  1    &  21    \\
f$_{\rm 6}$ & 191.027 (25) &  5234.87  (69)    &  0.874 (98)  &   7.11   &  1    &  20    \\
f$_{\rm 7^{\dagger\dagger}}$  & 237.895 (32)	&  4203.52	(57)    &	0.794	(10)	&  6.10  & 1 & 16 \\ 
\noalign{\smallskip}
\hline                                     
\label{138707823}
\end{tabular}
\end{table}

\subsubsection{TIC\,415339307}

\begin{figure*}
    \centering    
    \includegraphics[clip,width=1.\linewidth]{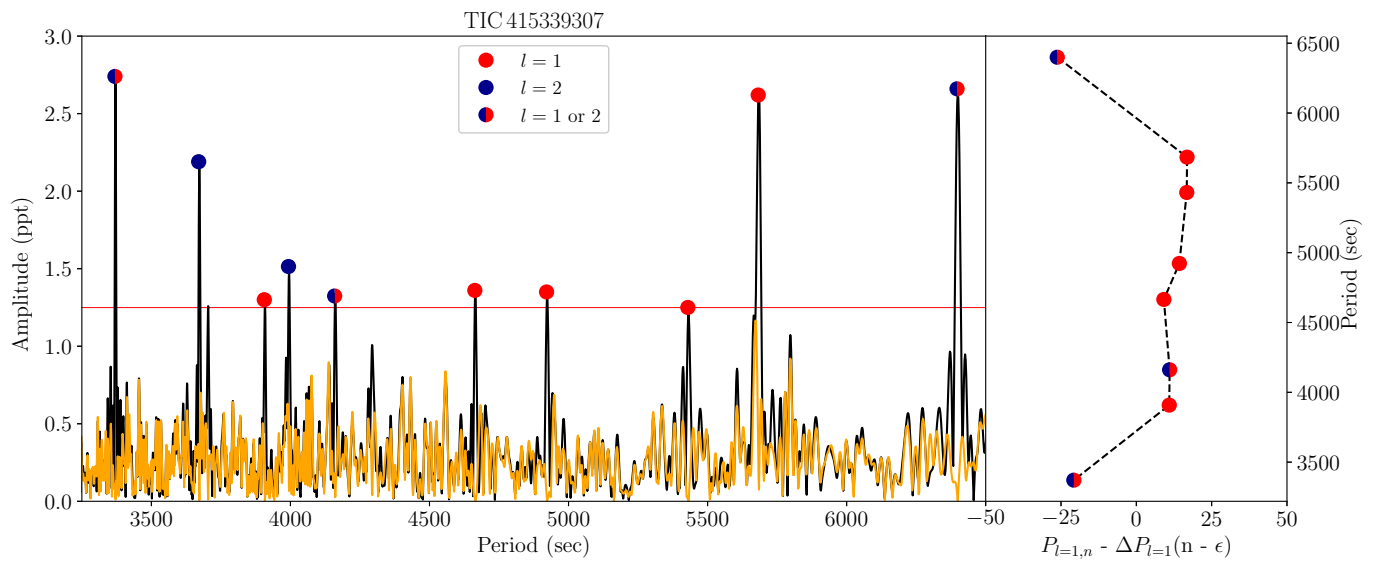} 
 \caption{Same as Fig. \ref{fig:260795163} but for TIC\,415339307 (presented here is sector 5).}
   \label{fig:415339307}
\end{figure*}

The FT of TIC\,415339307 shows that all the frequencies are concentrated in a narrow region between 156\,$\mu$Hz (3\,371\,s) and 296\,$\mu$Hz (6\,399\,s) as can be seen in Fig. \ref{fig:415339307}.   
In Fig. \ref{fig:ks-iv}, K-S and I-V tests show a sign of a potential average period spacing at 250\,s. 
We identified 5 dipole modes and 2 quadrupole modes, while 3 modes could fit both solutions. 
Between 4\,500\,s and 6\,000\,s, there are four frequencies that can only be fitted by dipole mode sequence. We find that the frequencies f$_{\rm 1}$, f$_{\rm 6}$, f$_{\rm 8}$ and f$_{\rm 10}$ also fit the dipole sequence.  
The frequencies f$_{\rm 7}$ and f$_{\rm 9}$ can only be quadrupole modes as the difference between the modes is around 300 s. 
The frequencies f$_{\rm 1}$, f$_{\rm 6}$ and f$_{\rm 10}$ could be identified as $l = 2$ modes.  

Including all these identified modes, we calculated $\Delta P_{l=1}$ = $252.84^{+3.30}_{-2.86}$ s and $\Delta P_{l=2}$ = $151.50^{+6.69}_{-2.06}$ s.  
The final seismic result for TIC\,415339307 are given in Table \ref{Table Seismic}.
We list all identified modes in Table \ref{415339307} and show them in Fig. \ref{fig:415339307}.

\begin{table}
\setlength{\tabcolsep}{2pt}
\renewcommand{\arraystretch}{1.1}
\centering
\caption{Frequency solution from the \emph{TESS} light curve of TIC\,415339307 including frequencies, periods, and 
amplitudes (and their uncertainties) and the signal-to-noise ratio. Identified modal degree and relative radial orders are listed column 5 and 6, respectively.}
\begin{tabular}{cccccccr}
\hline
\noalign{\smallskip}
ID & Frequency & Period & Amplitude & S/N & $l$ & $n$  \\
   & $\mu$Hz   &  [sec] &  [ppt]    &     &     &       \\
\noalign{\smallskip}
\hline
\noalign{\smallskip} 
f$_{\rm 1}$ &  156.261  (20)   &     6399.55 (79)    &    2.661 (21)   &    9.89   &   1/2 & 24/41     \\
f$_{\rm 2}$ &  175.929  (19)   &     5684.11 (63)    &    2.619 (21)   &    9.74   &   1   & 21        \\
f$_{\rm 3}$ &  184.122  (44)   &     5431.18 (1.26)  &    1.198 (21)   &    4.45   &   1   & 20        \\
f$_{\rm 4}$ &  203.126  (38)   &     4923.05 (92)    &    1.346 (21)   &    5.00   &   1   & 18        \\
f$_{\rm 5}$ &  214.359  (37)   &     4665.07 (82)    &    1.362 (21)   &    5.06   &   1   & 17        \\
f$_{\rm 6}$ &  240.309  (36)   &     4161.31 (67)    &    1.324 (21)   &    4.92   &   1/2 & 15/26     \\
f$_{\rm 7}$ &  250.302  (33)   &     3995.16 (54)    &    1.514 (21)   &    5.63   &   2   & 25        \\
f$_{\rm 8}$ &  255.863  (34)   &     3908.34 (60)    &    1.3   (21)   &    4.83   &   1   & 14        \\
f$_{\rm 9}$ & 272.299  (23)   &     3672.43 (56)    &    2.188 (21)   &    8.13   &   2   & 23        \\
f$_{\rm 10}$ & 296.648  (18)   &     3371.00 (21)    &    2.736 (21)   &    10.17  &   1/2 & 12/21     \\  
\noalign{\smallskip}
\hline  
\label{415339307}
\end{tabular}
\end{table}

\newcommand{\bftab}{\fontseries{b}\selectfont}

\begin{table}
\centering
\renewcommand{\arraystretch}{1.5}
\setlength{\tabcolsep}{2.1pt}
\caption{The seismic properties of 5 pulsating sdB stars studied in this work. Columns 1, 2, 3, 4, 5, 6 and 7 correspond to the {\it TESS} input catalog number, the number of g-modes extracted from the light curve, the number of identified dipole modes, the number of identified quadrupole modes, the range of radial orders, the average period spacing of dipole modes and the average period spacing of quadrupole modes, respectively.}
\begin{tabular}{cccccccc}
\hline 
TIC  & \# g & $l = 1$ & $l = 2$ & $ n $ & $\Delta P_{l=1}$ & $\Delta P_{l=2}$ \\
\noalign{\smallskip}
\hline
\noalign{\smallskip}                           
260795163 & 23 & 11 & 7 & 14-44  & $254.83^{+2.14}_{-2.28}$ & $150.51^{+3.27}_{-2.28}$ \\
080290366 & 18 & 11 & 3 & 14-48  & $253.32^{+0.78}_{-0.84}$ & $154.16^{+2.47}_{-7.97}$ \\
020448010 & 15 & 14 & 1 & 20-56  & $251.70^{+0.87}_{-0.96}$ & 145.32 \\
138707823 &  7 & 7 & - & 16-45   & $256.09^{+6.97}_{-1.21}$ & 147.83 \\
415339307 & 10 & 8 & 2 & 12-41   & $252.84^{+3.30}_{-2.86}$ & $151.50^{+6.69}_{-2.06}$ \\
\noalign{\smallskip}
\hline                                      
\label{Table Seismic}
\end{tabular}
\end{table}

\section{Analysis of spectroscopic data}
\label{sect:spect_data}

The data from the EFOSC2 spectrograph were reduced and analyzed using standard  \textit{PyRAF}\footnote{\url{http://www.stsci.edu/institute/software_hardware/pyraf}} \citep{pyraf2012} procedures. First, bias correction and flat-field correction have been applied. Then, the pixel-to-pixel sensitivity variations were removed by dividing each pixel with the response function. 
After this 
we applied wavelength calibrations using the spectra obtained with the internal He-Ar comparison lamp. 
In a last step, flux calibration was applied using the standard stars EG\,21 and EG\,274. 
The SNR of the final spectra is between 80 and 150 (see Table \ref{tablespec1}).

The data obtained with the B\&C spectrograph were reduced in similar way using \textit{PyRAF}. 
All frames were bias subtracted, flat-field corrected and cosmic ray events were removed.  
Afterwards, the wavelength calibration was performed with calibration spectra taken immediately after target observations. 
All wavelength-calibrated spectra were corrected for atmospheric extinction using coefficients provided by \textit{PyRAF}. 
Finally, all spectra were flux calibrated using the spectrophotometric standard star EG\,21. 
The final spectra have SNR, that is ranging from 70 to 120. 

\section{Spectral analysis with {\sc XTgrid}}
\label{sect:spect_analysis}

All stars have been analyzed and atmospheric parameters were derived by fitting synthetic spectra to the newly obtained low-resolution spectra. 
Synthetic spectra have been calculated from {\sc Tlusty} non-Local Thermodynamic Equilibrium stellar atmosphere models \citep{hubeny2017} using H and He composition. 
These models were utilized in the steepest-descent $\chi^2$ minimizing fitting procedure {\sc XTgrid} \citep{nemeth2012} using the web-service provided by Astroserver\footnote{\url{https://xtgrid.astroserver.org}}. 
The iterative procedure starts out from a starting model and by successive corrections converge on the best-fit. 
We applied a convergence limit of 0.5\% relative change of all model parameters over three successive iterations. 
Error bars were calculated by mapping the $\chi^2$ landscape around the best fit until the 3$\sigma$ confidence limit for the given degree-of-freedom was reached.
The error calculations are performed in one dimension for the He abundance, but include the correlations between surface temperature and gravity.

\begin{figure*}
    \centering
    \includegraphics[width=\textwidth]{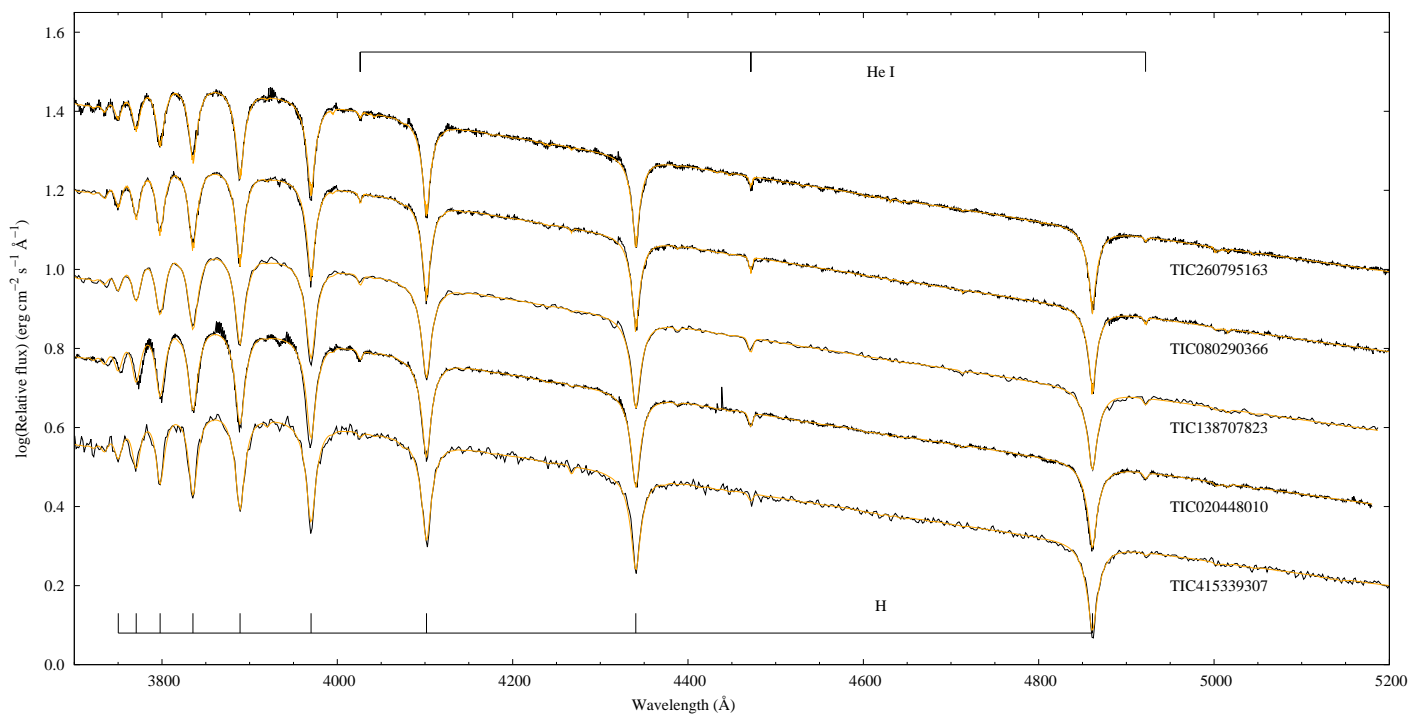}
     \caption{Best fit {\sc Tlusty/XTgrid} models for the five hot subdwarf stars analysed in this paper.
     The spectra are dominated by the H Balmer-series marked at the bottom and few weak He I lines marked at the top of the figure.
    The observed continua have been adjusted to the theoretical models.
    }
    \label{fig:xtgrid}
\end{figure*}

Figure \ref{fig:xtgrid} shows the new observations together with their best-fit {\sc Tlusty/XTgrid} models and Table \ref{tablesresult} lists the atmospheric parameters. 
The sample is very homogeneous, all spectra in Figure \ref{fig:xtgrid} are dominated by Balmer-lines and only weak He\,{\sc I} lines are seen. 
This, together with the Balmer-decrement suggest, that the stars must have very similar atmospheric parameters, which lines up with earlier observations, that sdBVs stars (g-mode pulsators) form a compact group on the EHB.
However, the spectral analysis revealed a systematic difference between the parameters derived from EFOSC2 and B\&C data. 
The poor blue coverage of the NTT/EFOSC2 wavelength calibration lamp results in serious flexure along the dispersion axis and cause discrepancies in fitting. 
For completeness we include all results in Table \ref{tablesresult}. 
Where B\&C spectra are available, we consider them superior in quality and the results from B\&C data as final.


Among the stars analyzed in this paper, there is a confirmed binary system TIC\,138707823. The spectrum of TIC\,138707823 in Figure \ref{fig:xtgrid} shows a clean sdB spectrum without any significant optical contribution from a companion. 
This confirms earlier results \citep{Edelmann2005, Geier2012}, that the companion is a compact object, most likely a WD.

\begin{table*}
\begin{center}
\caption{Spectroscopic results of five sdBV stars analyzed in this paper. 
The errors are 3$\sigma$ statistical errors.
Systematic differences exist between parameters from different instruments.
Where available, B\&C observations are adopted. 
The final parameters are given with bold TIC numbers.}
\begin{tabular}{ccccccc}
\hline \hline
TIC &  Spectrograph & Sp. Type & $T_{\rm eff}$ (K) & $\log{g}$ (cm\,s$^{-2}$) & $\log{(n_{\rm He} / n_{\rm H})}$ \\
\hline
\setrow{\bfseries}260795163  &B\&C  & sdB    & 27600 ($\pm$460)  & 5.37 ($\pm$0.08) & -2.76 ($\pm$0.02) \\
260795163                    &EFOSC2& sdB    & 26260 ($\pm$550)  & 5.12 ($\pm$0.09) & -2.80 ($\pm$0.09) \vspace{2pt}\\
\setrow{\bfseries}080290366  &B\&C  & sdB    & 26310 ($\pm$410)  & 5.30 ($\pm$0.05) & -2.61 ($\pm$0.08) \\
080290366                    &EFOSC2& sdB    & 25770 ($\pm$380)  & 5.21 ($\pm$0.06) & -2.69 ($\pm$0.06) \vspace{2pt}\\
\setrow{\bfseries}138707823  &B\&C  & sdB+WD & 24800 ($\pm$310)  & 5.49 ($\pm$0.06) & -2.57 ($\pm$0.05) \\
138707823                    &EFOSC2& sdB+WD & 26310 ($\pm$330)  & 5.37 ($\pm$0.08) & -2.53 ($\pm$0.07) \vspace{2pt}\\
\setrow{\bfseries}020448010  &EFOSC2& sdB    & 23720 ($\pm$480)  & 5.45 ($\pm$0.08) & -2.57 ($\pm$0.52) \vspace{2pt}\\
\setrow{\bfseries}415339307  &B\&C  & sdB    & 25920 ($\pm$530)  & 5.44 ($\pm$0.10) & -3.00 ($\pm$0.03) \\
\hline 
\label{tablesresult}
\end{tabular}
\end{center}
\end{table*}

\section{Asteroseismic models}
\label{sect:model_seismology}
\begin{figure*}
    \centering    
    \includegraphics[width=1\textwidth]{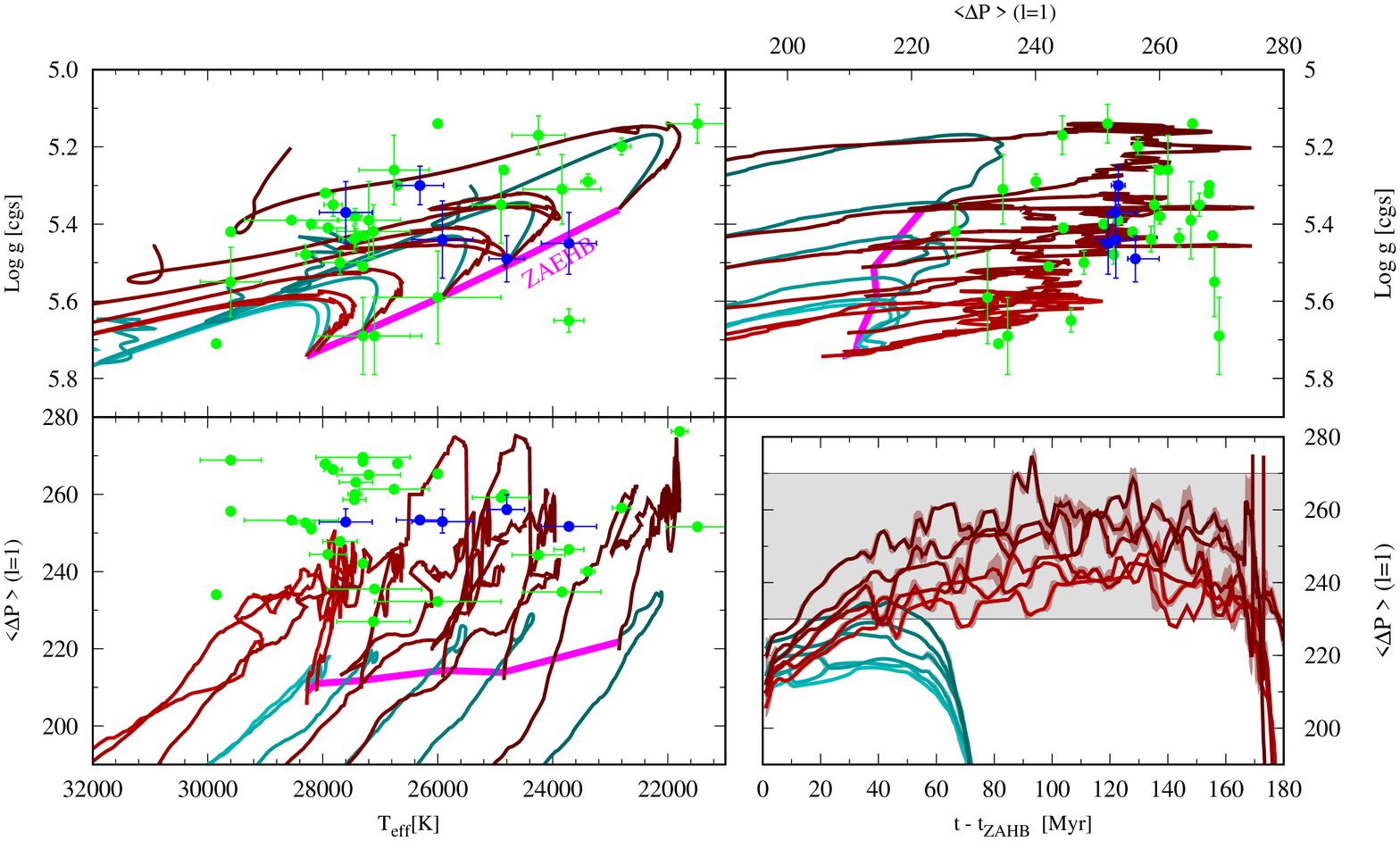} 
 \caption{Properties of the two sets of stellar evolution models discussed in the text as compared with those observed in pulsating sdB stars. Blue circles indicate the stars studied in this work while green circles indicate those from previous studies \citep{Reed2018,charpinet2019, reed2020, saho2020}. Red tracks correspond to models computed with a moderate CBM prescription while cyan tracks correspond to those computed under the extreme assumption of a strict Schwarzschild criterion. Masses of the models are $M_{\rm ZAHB}=0.467$, 0.4675, 0.468, and 0.47 $M_\odot$, darker colors correspond to more massive models. Thick magenta lines show the locus of model at the Zero Age Extreme Horizontal Branch (ZAEHB). Light coloured bands around the curves in the bottom right panel indicate the typical variance of $\Delta P$ around the mean value.}  
    \label{fig:models}
\end{figure*}

A proper asteroseismic interpretation of the observed frequencies found in g-mode sdB pulsators requires the computation of oscillations in stellar models \citep{charpinet2000,charpinet2002}. To this end we have computed stellar evolution models with {\tt LPCODE} stellar evolution code \citep{althaus2005}, and computed $l = 1$ g-mode frequencies with the adiabatic non-radial pulsation code {\tt LP-PUL} \citep{Corsico2006a}. Opacities, nuclear reaction rates, thermal neutrino emission and equation of state are adopted as in \cite{millerbertolami2016} and \cite{moehler2019}. In particular, it is worth noting that atomic diffusion was not included in the present computations. Diffusion is expected to reduce mode-trapping features in the latter stages of the He-core burning phase \citep{MillerBertolami2012}. In the present work we aim at the global comparison of the model predictions with observations by means of a robust indicator such as the mean period spacing ($\Delta P$), leaving detailed period-to-period comparisons of individual stars for future works. As such, we expect trapping features to play no important role in the comparisons. SdB models were constructed from an initially $M_i=1 M_\odot$ ZAMS model with $Z_{\rm ZAMS}=0.02$, and $Y=0.245+ 2 \times Z$, as in \cite{millerbertolami2016}. Mass loss was artificially enhanced prior to the He-core flash in order to produce sdB models during the core He burning stage in the $T_{\rm eff}$ range of g-mode sdB pulsators. The resulting models at the beginning of the He-core burning stage have masses of $M_{\rm ZAHB}=0.46738$, 0.4675, 0.468, 0.469, 0.47, and 0.473 $M_\odot$. For the sake of comparison, models during the He-core burning stage were computed under two different assumptions regarding convective boundary mixing (CBM). One is the extreme assumption of a strict Schwarzschild criterion \citep{schwarzschild1906} at the convective core, and the other corresponds to the inclusion of CBM at the boundary of the convective core. In the latter case, CBM is adopted as an exponentially decaying velocity field following \cite{freytag1996} and \cite{herwig1997}, with free parameters taken as in \cite{millerbertolami2016}. This corresponds to the assumption of a moderate CBM \citep[see section 3.1 in][]{degeronimo2019}. It is worth noting that although strong theoretical arguments indicate that strict Schwarzschild criterion is not physically sound, it serves as a useful estimation of the smallest possible convective core size \citep{castellani1971, castellani1985, gabriel2014}. The value of $\Delta P$ was computed from the periods of individual $l = 1$ g-modes in the range 2000\,s\textemdash 10000\,s, which is a typical range for the periods observed in V1093 Her stars. The computed value of $\Delta P$ shows oscillations due to two completely different reasons.  On one hand, as the models evolve structural changes push different periods inside, or outside, the range in which we computed $\Delta P$, leading to small fluctuations in the value of $\Delta P$. For the sake of clarity in Fig. \ref{fig:models} we show its time-averaged value in a moving window of 3 Myr. The variance around the mean value is shown by the confidence bands in the bottom right panel of Fig. \ref{fig:models}. On the other hand, larger fluctuations occurring on longer timescales also appear in the sequences that include moderate CBM due to the oscillating behaviour of the boundary of the convective core.

The two sets of sdB models are shown in Fig. \ref{fig:models} together with the properties observed in the V1093 Her stars studied in the present and previous works. From Fig. \ref{fig:models} it becomes apparent that models with a small convective core, close to those predicted by a strict Schwarzschild criterion, are too compact (and consequently too dim) to fit the surface gravities observed in known pulsators. This is reinforced by the comparison between the evolution of the mean period spacing ($\overline{\Delta P_\ell}$) against $\log g$ and $\log T_{\rm eff}$ in Fig. \ref{fig:models}. Models with small convective cores give mean period spacings too small to account for the observations. On the contrary, models with a moderate CMB prescription are able to reach the range of mean period spacings observed in V1093 Her stars (Fig. \ref{fig:models}). The reason behind this is two fold, and can be understood by looking at the expression of the asymptotic mean period spacing $\Delta P_\ell ^a$,
\begin{equation}
    \Delta P_\ell^a = \frac{P_o}{\sqrt{\ell(\ell+1)}}=
    \frac{2\pi^2}{\sqrt{\ell(\ell+1)}}\left[\int_{r_1}^{r_2} \frac{N}{r} dr\right]^{-1}.
    \label{eq:asymptotic_N}
\end{equation}
The buoyancy (Brunt-V\"ais\"al\"a) frequency for an ideal monoatomic gas with radiation can be written as 
\begin{equation}
    N^2=g^2\frac{\mu (4-3\beta)}{\Re\, T}
    \left[\nabla_{\rm ad}-\nabla+\frac{\beta\,\nabla_\mu}{4-3\beta}\right],
\end{equation}
where in the case of sdB stars radiation pressure is almost negligible and $\beta\sim 1$. We see that period spacing, in general, scales as $\Delta P\propto\, {\overline{g}}^{-1} R_\star/(R_\star-R_{\rm core})$. As models that include CBM have cores that grow larger than those using a strict Schwarzschild criterion the g-mode propagation cavity (of size $\sim R_\star-R_{\rm core}$) is reduced and leads to larger values of the period spacings. This is in fact the reason why models that include CBM show oscillations in the value of $\Delta P$, as those models show oscillations in the size of the convective core (see Fig. \ref{fig:models}). In addition, models with larger convective cores are able to increase the mean molecular weight of the star to larger values than their small convective core counterparts. This leads to larger luminosities in the sdB models that include CBM, and consequently to smaller surface gravities for the same effective temperature. The decrease in the mean value of the local surface gravity then also leads to an increase in the period spacing of the models. Finally, close to the end of the He-core burning stage, sdB models with CBM develop one of more breathing pulse instabilities \citep{castellani1985}. This creates the loops in the $\log g - \log T_{\rm eff}$ diagram at relatively low gravities and leads to an extension of the He-core burning lifetime.

From the previous discussion it is apparent that small convective cores close to those predicted by a bare Schwarzschild criterion can be discarded on the base of the observations. This is in agreement with theoretical arguments \citep{castellani1985, gabriel2014} as well as with independent observational constraints \citep{charpinet2011,  bossini2015, constantino2015, constantino2016}. Although models with a moderate CBM prescription at the burning core cover the range of observed period spacings, a closer inspection shows that observed mean period spacings are about 10 to 20\,s larger ($\sim 5$ to 10\%) than those predicted by the models. Such a shift could be attained by a small increase in the size of the convective core. The consequent small decrease in the surface gravity of the models would still be in agreement with observations.

\section{Summary and conclusions}
\label{sect:summary}

We present here the analysis of data collected on 5 pulsating hot subdwarf B stars  TIC\,260795163, TIC\,080290366, TIC\,020448010, TIC\,138707823 and TIC\,415339307 observed with the \textit{TESS} mission.  
From the five analyzed stars four are new detections of long-period pulsating sdB (V1093\,Her) stars, namely TIC\,260795163, TIC\,020448010, TIC\,138707823 and TIC\,415339307. 
This high-duty cycle space photometry delivered by the \textit{TESS} mission provides data with excellent quality to detect and identify the modes in long-period sdBV stars. 

The pulsations detected in these 5 stars are concentrated in the short frequency region from $\sim$70 to $\sim$300 $\mu$Hz, which is in line with what has been discovered during the second half of the survey phase of \textit{Kepler} \citep{baran2011}. We have detected 73 oscillation frequencies which we associate with g-modes in sdBVs.
We did not find any p-modes for any of the targets.
With the 120-sec observations, it is difficult to find p-modes since we are limited to Nyquist frequency of about 4\,200 $\mu$Hz. However, we did not find any p-modes either in the 20-sec data of TIC\,260795163, TIC\,080290366 and TIC\,138707823 even though the Nyquist frequency in that data set is about 25\,000 $\mu$Hz. This might imply that the analyzed stars are most likely pure g-mode sdB pulsators.
 
We have analyzed the data using the asteroseismic methods of rotational multiplets and asymptotic period spacing in order to identify pulsational modes. 
Although we detect many pulsation frequencies, we did not find evidence for complete rotational multiplets in any of the analyzed stars. 
Relying solely on asymptotic period spacing relationships, we identify the observed periods as mainly dipole and quadrupole g-modes. 

We constructed sFTs to examine the temporal evolution of the pulsation modes. 
The pulsation spectra and sFTs of all analyzed stars do not show clear pattern indicating rotational multiplets. Moreover, the highest signal-to-noise frequencies seem to be stable over the course of these \textit{TESS} observations (from 25 to 49 days).
However, the low-amplitude frequencies do show complex pattern in their sFTs. 
With a higher number of data points and with a longer baseline such as for stars that have been observed in multiple sectors 
with \textit{TESS}, a better frequency resolution will be achieved and also noise level will be lower, which will make detection of possible rotational multiples viable.
In particular, we believe that for the stars which lie in the continuing viewing zone of \textit{TESS} it would be possible to detect rotational multiplets and make the mode identification more secure. 
We will concentrate on the stars which have been observed in more than a single sector, and on a star which is in the southern continuous viewing zone of \textit{TESS} in forthcoming papers.

In the 5 V1093\,Her stars analyzed in this paper TIC\,260795163, TIC\,080290366, TIC,020448010 TIC\,138707823 and TIC\,415339307, we have identified 49 frequencies out of 73 detected ones as $l = 1$ modes by using solely asymptotic period spacing. 
In some cases we were not able to identify the modes with a unique modal degree. We mention that for 15 periodicities, we did not find a unique identification, therefore they can be interpreted as either $l = 1$ or $l = 2$ modes.

We apply the method of bootstrapping with $10^{4}$ times randomization in order to calculate the mean  period  spacing obtained in our analysis and also to assess the errors. The mean period spacing for $l = 1$ modes obtained in this way for the 5 analyzed stars is ranging from 251\,s to 255\,s, while the average period spacing for $l = 2$ modes is spanning from 145\,s to 155\,s. 
 
We have derived atmospheric parameters for all 5 stars analyzed in this paper by fitting synthetic spectra to the newly obtained low-resolution Dupont/B\&C and NTT/EFOSC2 spectra.
The effective temperature of the observed sdB stars is spanning from 23,700\,K to 27\,600\,K and their surface gravity ($\log{g}$) is in the range from 5.3 to 5.5\,dex, confirming that they are indeed occupying the g-mode sdBV parameter space.

We have computed stellar evolution models with {\tt LPCODE} stellar evolution code \citep{althaus2005}, and we have computed $l = 1$ g-mode frequencies with the adiabatic non-radial pulsation code {\tt LP-PUL} \citep{Corsico2006a}.
We compared the derived mean period spacings ($\Delta P$) of dipole g-modes derived from \textit{TESS} observations of 5 sdBV stars analysed in this paper and 33 found in the literature with the predictions of the adiabatic pulsation computations performed on stellar evolutionary models. 
In agreement with the expectations from theoretical arguments and previous asteroseismological works \citep{castellani1985, gabriel2014, charpinet2011, bossini2015, constantino2015, schindler2015}, and recently \citep{Ostrowski2020}, we find that models relying on a simplistic implementation of the Schwazschild criterion leads to small convective cores and values of $\Delta P$ which are too low to match the observations. On the contrary, models with standard treatment of convective boundary mixing at the convective core are able to match the observed values of $\Delta P$, although more intense convective boundary mixing cannot be discarded.

For any of the targets analysed in this work, we did not attempt to constrain envelope properties because of the following reasons. First, we did not detect any p-modes. Although lower order g-modes are sensitive to envelope properties, one needs to attempt to do a model fit of the observed periods as it has been done by \citep{charpinet2019}, which is beyond the scope of this paper. Lastly, we did not detect any trapped modes either, which can be used to investigate the envelope by identifying the period spacing between the trapped modes.

 
 
\begin{acknowledgements}
We  wish  to  acknowledge  the  suggestions  and comments of an anonymous referee that improved the original version of this work.
We thank Alejandro H. C\'orsico for kindly providing us with the codes for the K-S and I-V tests and for his valuable comments on the earlier version of this manuscript. 
M.U. thanks \"Ozg\"ur Bast\"urk and M.Ekrem Esmer for valuable discussions.
M.U. acknowledges financial support from CONICYT Doctorado Nacional in the form of grant number No: 21190886 and ESO studentship program.
P.N. acknowledges support from the Grant Agency of the Czech Republic (GA\v{C}R 18-20083S). 
This research has used the Sandbox services of \mbox{\url{www.Astroserver.org}}.
Financial support from the Polish National Science Centre under projects No.\,UMO-2017/26/E/ST9/00703 and UMO-2017/25/B ST9/02218 is acknowledged.
This paper includes data collected by the \textit{TESS} mission. Funding for the \textit{TESS} mission is provided by the NASA Explorer Program. Funding for the \textit{TESS} Asteroseismic
Science Operations Centre is provided by the Danish National Research Foundation (Grant agreement no.: DNRF106), ESA PRODEX (PEA 4000119301) and Stellar Astrophysics Centre (SAC) at Aarhus
University. We thank the \textit{TESS} team and staff and TASC/TASOC for their support of the present work.
We thank Brad Barlow, who made possible \textit{TESS} Cycle 3 observations of variable hot subdwarf stars observations with the proposal number G03221.
And finally we thank the TASC WG8 team for supporting this project and providing valuable feedback.

\end{acknowledgements}

\bibliographystyle{aa}
\bibliography{myrefs.bib}

\end{document}